# Magnetic bag like solutions to the SU(2) monopole equations on $\mathbb{R}^3$

Clifford Henry Taubes[†]

Department of Mathematics
Harvard University
Cambridge, MA 02138

(chtaubes@math.harvard.edu)

Abstract: Solutions to the SU(2) monopole equations in the Bogolmony limit are constructed that look very much like Bolognesi's conjectured magnetic bag solutions. Three theorems are also stated and proved that give bounds in terms of the topological charge for the radii of balls where the solution's Higgs field has very small norm.

[†] Supported in part by the National Science Foundation.

# 1. Introduction

The purpose of this note is to construct smooth solutions to the SU(2) Bogolmony equations that have properties that are attributed to what are called magnetic bags. There is no formal definition of a magnetic bag solution, but there is a list of desired properties, see [Bol], [LW], [M], [HPS] and [EG]. In short, the Higgs field of a magnetic bag solution must be nearly zero in a ball whose radius is on the order of topological charge and the solution must be nearly Abelian on the complement of a shell at this radius with thickness on the order of the square root of the topological charge. The upcoming Theorem 1.4 supplies solutions with large topological charge that come very close to having these properties. Theorems 1.1, 1.2 and 1.3 give an answer of sorts to a conjecture about the minimum radius ball in $\mathbb{R}^3$ that contains the region where the Higgs field is small. Analogs of Theorems 1.1-1.4 that concern Manton's conjectured 'nested bag' configurations [M] can likely be proved with the techniques that prove Theorems 1.1-1.4. By way of a parenthetical remark, solutions with the Yang-Mills-Higgs equations on $\mathbb{R}^3$ with infinite energy were discussed in the physics literature many years ago. See [S1], [S2] and the references in these papers.

The promised Theorems 1.1-1.4 are in Part 4 of this introduction. Parts 1, 2 and 3 set the stage and the notation for Part 4.

*Part 1*: The Bogolmony equations are a system of partial differential equations for a pair $(A, \Phi)$ with A being a connection on the product principle SU(2) bundle over $\mathbb{R}^3$ and with $\Phi$ being a section of the associated vector bundle with fiber the Lie algebra of SU(2). The latter is also the product bundle. These equations ask that the curvature 2-form of A, the A-covariant derivative of $\Phi$ and the pointwise norm of $\Phi$ obey

$$*F_A = d_A \Phi \quad and \quad \lim_{|x|\to\infty} |\Phi| = 1 \ . \tag{1.1}$$

with the notation such that $F_A$ denotes the curvature of A, and $d_A\Phi$ denotes the A-covariant derivative of $\Phi$. Meanwhile, $*$ denotes the Euclidean metric's Hodge star and the norm $|\cdot|$ is defined as follows: Use $\mathfrak{su}(2)$ in what follows to denote the Lie algebra of SU(2), this being the vector space of $2 \times 2$ traceless, anti-Hermitian complex matrices. An ad(SU(2)) invariant inner product on $\mathfrak{su}(2)$ assigns to a given pair $\mathfrak{a}, \mathfrak{b}$ the number -2 trace($\mathfrak{ab}$). This inner product is denoted by $\langle \mathfrak{ab} \rangle$. It is positive definite and $|\cdot|$ is the associated norm.

*Part 2*: What is done in the author's Ph.D. thesis (see Theorems 10.3 and 10.5 in Chapter IV in [JT]) implies that any pair $(A, \Phi)$ of connection on the product principal

SU(2) bundle and section of the product $\mathfrak{su}(2)$ bundle with both $|F_A|^2$ and $|d_A\Phi|^2$ being integrable functions on $\mathbb{R}^3$ has the following properties:

- *There exists $r \in [0, \infty)$ such that the 6'th power of $r - |\Phi|$ is integrable on $\mathbb{R}^3$.*
- *If $r > 0$, then $(4\pi r)^{-1} \int_{\mathbb{R}^3} \langle F_A \wedge d_A\Phi \rangle$ is an integer.*
- *If $r > 0$ and if there exists $R > 0$ such that $|\Phi| > 0$ where $|x| > R$, then this integer is the degree of the map that is defined by $\frac{\Phi}{|\Phi|}$ from sphere centered at the origin of radius greater than $R$ to the radius 1 sphere in $\mathfrak{su}(2)$.*

(1.2)

Now suppose that $(A, \Phi)$ is such that the norms of $F_A$ and $d_A\Phi$ are square integrable and such that $\lim_{|x|\to\infty} |\Phi| = r$. The third bullet in (1.2) asserts that $\frac{\Phi}{|\Phi|}$ defines a degree N map from every large radius sphere about the origin in $\mathbb{R}^3$ to the unit sphere in $\mathfrak{su}(2)$. As such, the $\mathfrak{su}(2)$ valued function $\Phi$ must vanish at points in $\mathbb{R}^3$ if N is not zero.

*Part 3*: Suppose that $(A, \Phi)$ is a solution to (1.1). As explained in Section 2d, both $|F_A|^2$ and $|d_A\Phi|^2$ are integrable functions on $\mathbb{R}^3$. It follows as a consequence that

$$\int_{\mathbb{R}^3} \langle F_A \wedge d_A\Phi \rangle = 4\pi N$$

(1.3)

with N being an integer. Note in particular that N is a priori non-negative because the equations in (1.1) imply directly that the integral on the left hand side of (1.3) is equal to $\frac{1}{4\pi}$ times the integrals over $\mathbb{R}^3$ of $|F_A|^2$ and $|d_A\Phi|^2$.

*Part 4*: Suppose that N is a positive integer and that $(A, \Phi)$ is a solution to (1.1) that obeys (1.3). As just noted, $|\Phi|$ must be zero on a non-empty set and thus small on some domain in $\mathbb{R}^3$. This fact begs the following questions:

- *What is the smallest radius of a ball in $\mathbb{R}^3$ that contains all points where $|\Phi| \ll 1$?*
- *What is the largest radius of a ball in $\mathbb{R}^3$ that contains only points where $|\Phi| \ll 1$?*

(1.4)

The three theorems that follow address these questions.

**Theorem 1.1**: *Suppose that N is a positive integer and that $(A, \Phi)$ is a solution to (1.1) that obeys (1.3). Fix $\varepsilon \in (\frac{1}{N}, 1)$ and let $R_\varepsilon$ denote the infimum of numbers $R \in (0, \infty)$ such that $|\Phi| > \varepsilon$ where $|x| > R$. Then $R_\varepsilon > N(1-\varepsilon)^{-1}$.*

**Theorem 1.2**: *Suppose that $N$ is a positive integer and that $(A, \Phi)$ is a solution to (1.1) that obeys (1.3). Fix $\varepsilon \in (0, 1)$ and let $r_\varepsilon$ denote the supremum of numbers $r \in [0, \infty)$ such that $|\Phi| < \varepsilon$ where $|x| < r$. Then $r_\varepsilon < N(1-\varepsilon)^{-1}$.*

**Theorem 1.3**: *Suppose that $N$ is a positive integer and that $(A, \Phi)$ is a solution to (1.1) that obeys (1.3). Fix $\varepsilon \in (0, 1)$ and let $\hat{r}_\varepsilon$ denote the supremum of numbers $r \in [0, \infty)$ such that $\frac{1}{4\pi r^2} \int_{|x|=r} |\Phi| < \varepsilon$. Then $\hat{r}_\varepsilon < N(1-\varepsilon)^{-2}$.*

Theorems 1.1-1.3 are proved in Section 2.

The fourth main theorem of this paper asserts the existence of solutions to (1.1) that obey (1.3) and come close to realizing the $\varepsilon = N^{-1/2}$ version of the lower bound in Theorem 1.1 and the $\varepsilon = N^{-1/2}$ version of the upper bound in Theorems 1.2 and 1.3.

**Theorem 1.4**: *There exists $\kappa > 1$ with the following significance: Suppose that $N$ is an integer that is greater than $\kappa$. There are solutions to (1.1) that obey (1.3) with the corresponding $\varepsilon = N^{-1/2}(\ln N)^{21}$ version of $R_\varepsilon$ at most $N(1+\kappa N^{-1/2}(\ln N)^{21})$ and with the corresponding $\varepsilon = \kappa^2 N^{-1/2}(\ln N)^{21}$ versions of $r_\varepsilon$ and $\hat{r}_\varepsilon$ no less than $N(1+\kappa N^{-1/2}(\ln N)^{21})$. Moreover, each zero of the $\Phi$ component of any such solution is in a spherical shell of thickness at most $\kappa N^{1/2}(\ln N)^{16}$ and inner radius sphere of radius $N(1+\kappa N^{-1/2}(\ln N)^{21})$.*

What is said in Theorem 1.4 implies that its solutions are very nearly magnetic bag solutions. Note in this regard that the arguments in the proof of Theorem 1.4 imply that its solutions looks like Abelian solutions up to an exponentially small term on the part of $\mathbb{R}^3$ where the distance to the radius $N(1+\kappa N^{-1/2}(\ln N)^{21})$ sphere about the origin is $\kappa N^{1/2}(\ln N)^{16}$ or more.

By way of a parenthetical remark, there is nothing significant about the exponent 21 for the power of lnN. The essential fact is that the construction of the solutions requires that some powers of lnN appear. Certain parameters were chosen for convenience that resulted in this power being 21. With more work, it is likely that Theorem 1.4 could be stated with a smaller power of lnN appearing.

Theorem 1.4 is proved in Section 4e assuming a technical proposition that is proved in Section 5. Section 3 and Sections 4a-d supply input for the proof.

## 2. The numbers $R_\varepsilon$ and $r_\varepsilon$

This section contains the proofs of Theorems 1.1, 1.2 and 1.3. The proofs are in reverse order. A final subsection ties up a loose end from Section 1 by proving the assertion in Part 3 of Section 1 to the effect that both $|F_A|^2$ and $|d_A\Phi|^2$ are integrable functions on $\mathbb{R}^3$ when $(A,\Phi)$ obeys both equations in (1.1).

### a) Proof of Theorem 1.3

Set $\delta_r$ to denote $\frac{1}{4\pi r^2} \int_{|x|=r} |\Phi|$. The fundamental theorem of calculus with the fact that $\lim_{|x|\to\infty} |\Phi| = 1$ implies that

$$1 - \delta_r \leq \int_r^\infty ds \left( \frac{1}{4\pi s^2} \int_{|x|=s} |d|\Phi|| \right) .$$

(2.1)

Since $|d|\Phi|| \leq |\nabla_A \Phi|$, the right hand side of (2.1) is no greater than

$$\int_r^\infty ds \frac{1}{\sqrt{4\pi}\, s} \left( \int_{|x|=s} |\nabla_A \Phi|^2 \right)^{1/2} ,$$

(2.2)

which is, in turn, at most the product of the square roots of two integrals, the first being the integral of the function $s \to \frac{1}{4\pi s^2}$ from $r$ to $\infty$, and the second being the integral of $|\nabla_A \Phi|^2$ on the $|x| \geq r$ part of $\mathbb{R}^3$. Since the integral of $|\nabla_A \Phi|^2$ over the whole of $\mathbb{R}^3$ is equal to $4\pi N$, it follows as a consequence that (2.2) is at most $r^{-1/2} N^{1/2}$. This being the case, what is said in (2.1) implies that

$$r \leq (1 - \delta_r)^{-2} N .$$

(2.3)

The assertion of Theorem 1.3 follows from (2.3) by taking $r$ so that $\delta_r = \varepsilon$.

### b) Proof of Theorem 1.2

The proof has three steps. The proof introduces by way of notation $d_A^\dagger$ to denote the formal, $L^2$ adjoint of the operator $d_A$; this is the operator that maps $\mathfrak{su}(2)$ valued 1-forms to $\mathfrak{su}(2)$ valued functions by the rule $a \to -*d_A *a$. Meanwhile, $d^\dagger$ is used in what follows to denote the formal adjoint of the exterior derivative on functions, this sending an $\mathbb{R}$ valued 1-form to a function by the rule $v \to -*d*$.

Step 1: The equations in (1.1) require that $\Phi$ obey the equation

$$d_A^\dagger d_A \Phi = 0 .$$

(2.4)

Take the inner product of both sides of this equation with $\Phi$ to see that $|\Phi|^2$ obeys

$$d^\dagger d |\Phi|^2 = -2|d_A\Phi|^2 .$$

(2.5)

This equality implies in turn that

$$d^\dagger d |\Phi| \leq 0 ,$$

(2.6)

this because the norm of $d|\Phi|$ is no greater than that of $d_A\Phi$.

    Step 2: The function on $\mathbb{R}^3$ that is defined by the rule

$$x \to 1 - \frac{r_\varepsilon}{|x|}(1-\varepsilon)$$

(2.7)

is harmonic and equal to $\varepsilon$ on the $|x| = r_\varepsilon$ sphere. Denote this function by g. This function is greater than $|\Phi|$ on the $|x| = r_\varepsilon$ sphere and both $|\Phi|$ and g to limit 1 as $|x| \to \infty$. These observations, (2.3) and the fact that g is harmonic imply via the maximum priniciple that $|\Phi| < g$ where $|x| > r_\varepsilon$.

    Step 3: Observations in the author's Ph. D. thesis (see Theorem 10.5 in Chapter IV of [JT]) imply that $|\Phi|$ where $|x|$ is large looks like $1 - \frac{N}{|x|} + \cdots$ where the unwritten terms limit to zero as $|x| \to \infty$ faster than $|x|^{-1}$. This depiction of $|\Phi|$ implies that $|\Phi|$ will be less than the function in (2.7) at large values of $|x|$ only if $r_\varepsilon$ is less than $N(1-\varepsilon)^{-1}$.

**c) Proof of Theorem 1.1**
    The proof has six steps.

    Step 1: Introduce by way of notation $\sigma$ to denote $\frac{\Phi}{|\Phi|}$, this being a map from the complement of the zeros of $\Phi$ to the unit sphere in $\mathfrak{su}(2)$. Use $\sigma$ to define the connection $\hat{A} = A - \frac{1}{4}[\sigma, d_A\sigma]$. The definition of $\hat{A}$ is such as to make $d_{\hat{A}}\sigma = 0$. The curvature of $\hat{A}$ is related to that of A by the formula

$$F_{\hat{A}} = F_A - \tfrac{1}{4} d_A\sigma \wedge d_A\sigma .$$

(2.8)

As the components of $F_{\hat{A}}$ commute with $\sigma$, so $F_{\hat{A}}$ can be written as $f\sigma$ with

$$f = \langle \sigma F_A \rangle - \tfrac{1}{4}\langle \sigma d_A\sigma \wedge d_A\sigma \rangle .$$

(2.9)

This is a closed 2-form that is defined on the complement of the zeros of $\Phi$.

Step 2: Fix $\varepsilon > 0$ and then fix $R > 0$ such that $|\Phi| \geq \varepsilon$ where $|x| \geq R$. What is said by the second bullet in (1.2), the fact that $d_A\Phi$ is square integrable, and the fact that $f$ is closed implies that the integral of $\frac{1}{4\pi} f$ over any constant $|x|$ sphere is equal to N if the radius of the sphere is R or greater. This is to say that

$$N = \frac{1}{4\pi} \int_{|x|=s} (\langle \sigma F_A \rangle - \tfrac{1}{4}\langle \sigma d_A\sigma \wedge d_A\sigma \rangle) \quad \text{if } s \geq R.$$

(2.10)

Multiply both sides of (2.10) by $s^{-2}$ and integrate the resulting identity between functions of s over the interval $[R, \infty)$ to see that

$$\frac{N}{R} = \frac{1}{4\pi} \int_R^\infty \frac{1}{s^2} \left( \int_{|x|=s} \langle \sigma F_A \rangle \right) ds - \frac{1}{16\pi} \int_R^\infty \frac{1}{s^2} \left( \int_{|x|=s} \langle \sigma d_A\sigma \wedge d_A\sigma \rangle \right) ds.$$

(2.11)

The next step supplies an upper bound for the left most integral on the right hand side of (2.11) and the subsequent two steps supply one for the right most integral.

Step 3: The left most term on the right hand side of (2.11) can be rewritten using (1.1) as

$$\frac{1}{4\pi} \int_R^\infty \frac{d}{ds}\left( \frac{1}{s^2} \int_{|x|=s} |\Phi| \right) ds.$$

(2.12)

Now use the fundamental theorem of calculus with the fact that $\lim_{|x|\to\infty} |\Phi| = 1$ to write the latter expression as

$$1 - \frac{1}{4\pi} R^{-2} \int_{|x|=R} |\Phi|.$$

(2.13)

Section 2a introduced $\delta_R$ to denote $\frac{1}{4\pi} R^{-2} \int_{|x|=R} |\Phi|$. Keep in mind that this $\delta_R$ is no less than $\varepsilon$ since $|\Phi| \geq \varepsilon$ on the $|x| = R$ sphere.

Step 4: The derivation of an upper bound for the right most integral in (2.11) starts with the observation that the latter is in no case greater than

$$\tfrac{1}{16\pi} \int_R^\infty \tfrac{1}{s^2} \Big( \int_{|x|=s} |d_A\sigma|^2 \Big)\, ds \;,$$

(2.14)

and thus no greater than

$$\tfrac{1}{16\pi} \tfrac{1}{R^2} \int_{|x|\geq R} |d_A\sigma|^2 \;.$$

(2.15)

Step 5 proves that the expression in (2.15) is at most $\tfrac{1}{4} \tfrac{N}{\varepsilon R^2}$.

<u>Step 5</u>: Let $\beta_\varepsilon$ denote the function on $[0, \infty)$ given by the rule $\beta_\varepsilon(t) = \varepsilon^{-1}t$ on $[0, \varepsilon]$ and given by the constant function 1 on $[\varepsilon, \infty)$. The salient features of $\beta_\varepsilon$ being that it equals 1 where $t > \varepsilon$, that it has non-negative derivative, and that it is proportional to t where t is near zero.

Define $\alpha_\varepsilon$ to be the function on $\mathbb{R}^3$ given by $\alpha = \beta(|\Phi|)$. Take the inner product of both sides of (2.4) with $\alpha_\varepsilon \sigma$ and use the fact that $|\sigma| = 1$ to derive the identity

$$-d^\dagger(\alpha_\varepsilon d|\Phi|) = \alpha_\varepsilon |\Phi|\, |d_A\sigma|^2 + (\beta_\varepsilon')_{t=|\Phi|} |d|\Phi||^2$$

(2.16)

Since $|\Phi| \geq \varepsilon$ where $|x| \geq R$, this equation leads to the inequality

$$|d_A\sigma|^2 \leq \varepsilon^{-1} \alpha_\varepsilon |\Phi|\, |d_A\sigma|^2 \;\; where\; |x| \geq R.$$

(2.17)

Use this inequality to see that

$$\int_{|x|\geq R} |d_A\sigma|^2 \leq \varepsilon^{-1} \int_{|x|\geq R} \alpha_\varepsilon |\Phi|\, |d_A\sigma|^2 \;.$$

(2.18)

As noted previously, the derivative of $\beta_\varepsilon$ is non-negative. It follows as a consequence that the side of right hand side of (2.18) is no greater than the integral over $\mathbb{R}^3$ of the right hand side of (2.16). This being the case, (2.18) leads to the inequality

$$\int_{|x|\geq R} |d_A\sigma|^2 \leq \varepsilon^{-1} \int_{\mathbb{R}^3} -d^\dagger(\alpha_\varepsilon d|\Phi|) \;.$$

(2.19)

Meanwhile, an integration by parts with the fact that $\alpha_\varepsilon = 1$ where $|\Phi|$ is nearly 1 can be used to see that the right hand side of (2.19) is equal to $\varepsilon^{-1} 4\pi N$.

<u>Step 6</u>: The identity in (2.11) with the bounds from Steps 3 and 4 imply that

$$\tfrac{N}{R} \le (1-\delta_R) + \tfrac{1}{4}\tfrac{N}{\varepsilon R^2} \;.$$

(2.21)

If $\varepsilon N > 100$, then this last inequality is obeyed when either

$$\tfrac{N}{R} \le (1-\delta_R) + \tfrac{1}{100}(\varepsilon N)^{-1} \quad or \quad \tfrac{N}{R} \ge 4\varepsilon N$$

(2.22)

As explained next, the right most inequality in (2.22) is not an option given the assumption that $\varepsilon N > 100$. To see why this is, note that the left most inequality demands that $R \ge N(1-\varepsilon + \tfrac{1}{100}(\varepsilon N)^{-1})^{-1}$ and the right most demands that $R \le \tfrac{1}{4}\varepsilon^{-1}$. If the right most inequality is obeyed for a given value of R, then $|\Phi| > \varepsilon$ on any $r > R$ sphere, this by assumption. It follows as a consequence that the right most inequality in (2.22) is not an option unless $\tfrac{1}{4}\varepsilon^{-1} \ge N(1-\varepsilon + \tfrac{1}{100}(\varepsilon N)^{-1})^{-1}$. This requires in turn that $\varepsilon N$ be less than $\tfrac{1}{2}$.

### d) The integrals of $|F_A|^2$ and $|d_A\Phi|^2$ when $(A, \Phi)$ obeys (1.1)

Suppose that $(A,\Phi)$ obeys both equations in (1.1). The six steps that follow of this subsection prove that the functions $|F_A|^2$ and $|d_A\Phi|^2$ are integrable on the whole of $\mathbb{R}^3$.

<u>Step 1</u>: Fix a favorite non-negative and non-increasing function on $\mathbb{R}$ that is equal to 1 on $(-\infty, \tfrac{1}{4}]$ and equal to zero on $[\tfrac{1}{2}, \infty)$. The chosen function is denoted in what follows by $\chi$. Given $R \in [1, \infty)$, use $\chi_{(R)}$ to denote the function $\mathbb{R}^3$ whose value at any given point x is $\chi(R^{-1}|x| - 1)$. This function is equal to 1 where $|x| < R$ and it is equal to zero where $|x| > 2R$. It is also that case that its derivative is bounded by an R-independent multiple of $R^{-1}$ and its second derivative where $|x| > 0$ is is bounded by an R-independent multiple of $R^{-2}$.

<u>Step 2</u>: The function $|\Phi|^2$ obeys (2.5) and this implies that its Laplacian is non-negative. This being the case, the maximum principal can be invoked to prove that $|\Phi|^2$ is no greater than 1 at any point. As a parenthetical remark, the strong maximum principle can be used to conclude that $|\Phi|^2$ is either strictly less than 1 at all points or else $d_A\Phi = 0$ at all points.

<u>Step 3</u>: Define the function $f$ on $[0, \infty)$ by the rule

$$r \to f(r) = \tfrac{1}{4\pi r^2} \int_{|x|=r} |\Phi|^2 \;.$$

(2.23)

It follows from (2.5) that this function obeys the differential inequality

$$\tfrac{d}{dr}(r^2 \tfrac{d}{dr} f) \geq 0 \; .$$

(2.24)

It follows as a consequence that $f$ obeys $f \geq 1 - \tfrac{\Delta}{r}$ with $\Delta$ being a positive constant. It follows from what is said in Step 2 that $f$ is nowhere greater than 1. This fact can also be seen from (2.23).

Step 4: Fix $R > 1$ and multiply both side (2.5) by $-\chi_{(R)}$. Then integrate the resulting equality over $\mathbb{R}^3$ to obtain an identity between integrals. Use integration by parts for the left hand integral in this identity to change the integrand $-\chi_{(R)} d^\dagger d |\Phi|^2$ to the integrand $(-d^\dagger d\chi_{(R)})(|\Phi|^2 - 1)$. There are no spurious boundary terms from this integration by parts because $\chi_R$ is zero where $|x| > 2R$.

Step 5: The right hand integral is the integral over $\mathbb{R}^3$ of $2\chi_{(R)} |d_A \Phi|^2$. This is no less than twice the integral of $|d_A \Phi|^2$ over the radius R ball centered at the origin. Meanwhile, it follows from what was said in Step 1 about the derivatives of $\chi_{(R)}$ and what is said in Step 2 about the positivity of $1 - |\Phi|^2$ that the integral of $(-d^\dagger d\chi_{(R)})(|\Phi|^2 - 1)$ is no greater than an R-independent constant times the integral over the $R \leq |x| \leq 2R$ shell of the function $R^{-2}(1 - |\Phi^2|)$. It follows from what is said in Step 3 that the latter integral is no greater than an R-independent constant times the number $\Delta$.

Step 6: It follows from what is said in Step 5 that there is an R-independent upper bound for the integral of $|d_A \Phi|^2$ over the radius R ball centered at the origin. This understood, an appeal to the Dominated Convergence Theorem (see for example [R]) proves that $|d_A \Phi|^2$ has finite integral over the whole of $\mathbb{R}^3$.

**3. A pair $(A, \Phi)$ that almost solves (1.1)**

Solutions to (1.1) and (1.3) that look like magnetic bags are constructed in Section 4 by a deforming a pair of connection on the product principal bundle and section of the product $\mathfrak{su}(2)$ bundle that obeys (1.3) and comes close to solving (1.1). This section constructs such a pair.

By way of a look ahead at the proof, the construction that follows of a pair that obeys (1.3) and comes close to solving (1.1) can be viewed as a modified version of the construction that was introduced in the author's Ph. D. thesis (see Chapter IV in [JT]) to construct solutions to (1.1) for any given N. The latter construction starts with N suitably chosen solutions to (1.1), each obeying

$$\int_{\mathbb{R}^3} \langle F_A \wedge d_A \Phi \rangle = 4\pi .$$

(3.1)

These are glued together with the result being a pair that obeys (1.3) and almost solves (1.1). This approximate solution to (1.1) is then perturbed to obtain an honest solution.

The construction used here replaces the N solutions to (1.1) with N solutions to only the left most equation in (1.1). Each pair in this new set obey $\lim_{|x| \to \infty} |\Phi| = r$ with $r$ being a suitable positive number. The integral on the left hand side of (3.1) for each pair is equal to $4\pi r$. Even so, each pair has the property that $\frac{\Phi}{|\Phi|}$ defines a degree 1 map from any sufficiently large, but constant $|x|$ sphere to the unit sphere in $\mathfrak{su}(2)$. The construction used in what follows glues these N solutions together so as to obtain a pair that obeys (1.3), is such that $\lim_{|x| \to \infty} |\Phi| = 1$, and very nearly obeys the equation $*F_A = d_A \Phi$.

**a) The Prasad-Sommerfeld solution**

The space of solutions to (1.1) that obeys (1.3) for a given integer N is denoted in what follows by $\mathcal{M}_N$. This space enjoys a free action of the group $C^\infty(\mathbb{R}^3; SU(2))$. It also enjoys an action of the group of isometries of $\mathbb{R}^3$. The quotient of $\mathcal{M}_0$ by $C^\infty(\mathbb{R}^3; SU(2))$ is a point, this being the equivalence class of the pair consisting of the product connection and a constant map to $\mathfrak{su}(2)$ with norm 1. The quotient of $\mathcal{M}_1$ by $C^\infty(\mathbb{R}^3; SU(2))$ is diffeomorphic to $\mathbb{R}^3$ via a diffeomorphism that intertwines the respective actions of the isometry group of $\mathbb{R}^3$. In particular, all N = 1 solutions to (1.1) are obtained from any given N = 1 solution by the action of elements in $C^\infty(\mathbb{R}^3; SU(2))$ and translations of $\mathbb{R}^3$. A particular N = 1 solution is known as the *Prasad-Sommerfeld* [PS] solution, this defined in the upcoming (3.2).

The definition of the Prasad-Sommerfeld solution refers to a basis, $\{\sigma_1, \sigma_2, \sigma_3\}$ for $\mathfrak{su}(2)$ with the property that each element has square equal to -1 times the identity matrix and such that $\sigma_1 \sigma_2 = -\sigma_3$. Note in particular that the set $\{\frac{1}{2} \sigma_k\}_{k=1,2,3}$ constitutes an orthonormal basis for $\mathfrak{su}(2)$. The definition uses $\theta_0$ to denote the product connection on the product principal SU(2) bundle $\mathbb{R}^3 \times SU(2)$. The connection component of the Prasad-Sommerfeld solution is denoted by $A_{PS}$ and written as $A_{PS} = \theta_0 + a_{PS}$ with $a_{PS}$ being an $\mathfrak{su}(2)$-valued 1-form on $\mathbb{R}^3$. The corresponding section of the product SU(2) bundle is denoted by $\Phi_{PS}$. The latter and $a_{PS}$ are given by the formula

$$a_{PS} = \left(\frac{1}{|x|} - \frac{1}{\sinh(|x|)}\right) \varepsilon_{ijk} \frac{x_i}{|x|} dx_j \tfrac{1}{2} \sigma_k \quad \text{and} \quad \Phi_{PS} = \left(\frac{1}{\tanh(|x|)} - \frac{1}{|x|}\right) \frac{x_i}{|x|} \tfrac{1}{2} \sigma_i$$

(3.2)

with it understood that the indices run over the set $\{1, 2, 3\}$, repeated indices are summed and with $\varepsilon_{ijk}$ being completely antisymmetric with $\varepsilon_{123} = 1$. The next paragraphs list some of the salient features of the pair $(A_{PS}, \Phi_{PS})$.

The first point of note is that $\Phi_{PS}$ has a single, non-degenerate zero, this at the origin. As can be seen directly from (3.2), $\Phi_{PS}$ near zero is $\frac{1}{6} x_i \sigma_i + \mathcal{O}(|x|^2)$. The remaining points concern $(A_{PS}, \Phi_{PS})$ where $|x| > 1$. The pair here differ from what is called the *Dirac* monopole by terms that are $\mathcal{O}(e^{-|x|})$. The Dirac monopole is a solution to (1.1) on $\mathbb{R}^3 - 0$, this denoted by $(A_D, \Phi_D)$ with the latter given by the formula

$$A_D = \theta_0 + \varepsilon_{ijk} \frac{x_i}{|x|^2} dx_j \tfrac{1}{2} \sigma_k \quad \text{and} \quad \Phi_D = (1 - \tfrac{1}{|x|}) \frac{x_i}{|x|} \tfrac{1}{2} \sigma_i .$$

(3.3)

As can be seen directly from (3.2) and (3.3), both $A_{PS} - A_D$ and $\Phi_{PS} - \Phi_D$ differ at points with $|x| > 1$ by at most $c_0 e^{-|x|}$. This is also the case for their derivatives. The next paragraph has more to say about the Dirac monopole.

Denote by $\sigma_D$ the map from $\mathbb{R}^3 - 0$ to the unit sphere about the origin in $\mathfrak{su}(2)$ that is defined by the rule $x \to \sigma_D(x) = \frac{x_i}{|x|} \tfrac{1}{2} \sigma_i$. This map is $A_D$-covariantly constant. Moreover, the curvature 2-form of $A_D$ is the $\sigma_D$ pull-back from the unit sphere in $\mathfrak{su}(2)$ of the constant, volume 2-form. To say more about this, let $(\theta, \varphi)$ denote the spherical coordinates for the $|x| = 1$ sphere, with $\theta$ giving the latitude and $\varphi$ giving the longitude. The pair $(A_D, \Phi_D)$ are such that

$$\langle F_{A_D} \sigma_D \rangle = \sin\theta \, d\theta \, d\varphi .$$

(3.4)

Rescaled and translated versions of $(A_{PS}, \Phi_{PS})$ are needed in what follows with each version defined by a some positive number, denoted here by $r$, and some point in $\mathbb{R}^3$, denoted here by p. The corresponding pair of connection on the product principal SU(2) bundle and section of product $\mathfrak{su}(2)$ bundle is given the formula

$$(\theta_0 + (\tfrac{1}{|x-p|} - \tfrac{r}{\sinh(r|x-p|)}) \varepsilon_{ijk} \tfrac{(x-p)_i}{|x-p|} dx_j \tfrac{1}{2} \sigma_k, (\tfrac{r}{\tanh(r|x-p|)} - \tfrac{1}{|x-p|}) \tfrac{(x-p)_i}{|x-p|} \tfrac{1}{2} \sigma_i) .$$

(3.5)

This pair obeys the left most equation in (1.1) but not the right most unless $r = 1$ because the $|x| \to \infty$ limit of $\tfrac{r}{\tanh(r|x-p|)}$ is $r$.

**b) Gluing solutions**

Fix $N > 1$. The three parts of this subsection construct a pair that obeys (1.3) and, for certain parameter choices, comes close to obeying (1.1). The pair that results from this construction is denoted by $(A_G, \Phi_G)$.

*Part 1*: Fix a configuration of N distinct points in $\mathbb{R}^3$. Use $\Theta$ to denote the chosen set of points. The constructions that follow require that the points in $\Theta$ obey certain constraints, the first set being the positivity of each $p \in \Theta$ version of the number

$$r_p = 1 - \sum_{q \in \Theta - p} \frac{1}{|p-q|} .$$

(3.6)

Fix a number to be denoted by L which is positive but no greater than half of the distance between the closest two points in $\Theta$. The notation in what follows uses $B_L(p)$ to denote the open, radius L ball with center on a given point $p \in \Theta$ and it uses $\bar{B}_{L/4}(p)$ to denote the closed, radius L/4 ball with center p. The balls from the collection $\{B_L(p)\}_{p \in \Theta}$ are pairwise disjoint because $L < \frac{1}{2}|p-q|$ if p and q are distinct points in $\Theta$.

A $C^\infty(\mathbb{R}^3; SU(2))$ equivalence class of a pair of connection on the product bundle principal bundle $\mathbb{R}^3 \times SU(2)$ principal bundle and section of $\mathbb{R}^3 \times \mathfrak{su}(2)$ is determined by the data listed next in (3.7) subject to the constraints listed in (3.8).

- *An $\mathfrak{su}(2)$-valued 1-form on $\mathbb{R} - (\cup_{p \in \Theta} \bar{B}_{L/4}(p))$ and map from $\mathbb{R}^3 - (\cup_{p \in \Theta} \bar{B}_{L/4}(p))$, these to be denoted by $a_\infty$ and $\Phi_\infty$.*
- *An assignment to each $p \in \Theta$ of an $\mathfrak{su}(2)$-valued 1-form on $B_L(p)$ and map from $B_L(p)$ to $\mathfrak{su}(2)$. These are denoted in what follows by $a_p$ and $\Phi_p$.*

(3.7)

The data set $\{(a_\infty, \phi_\infty), \{(a_p, \phi_p)\}_{p \in \Theta}\}$ must obey the following constraints:

*Fix $p \in \Theta$. There exists a map $g_p: B_L(p) \to SU(2)$ such that*
$a_p = g_p^{-1} a_\infty g_p + g_p^{-1} dg_p$ *and* $\phi_p = g_p^{-1} \phi_\infty g_p$ *on* $B_L(p) - \bar{B}_{L/4}(p)$.

(3.8)

By way of an explanation, the set $\mathbb{R}^3 - (\cup_{p \in \Theta} \bar{B}_{L/4}(p))$ with the balls that comprise the collection $\{B_L(p)\}_{p \in \Theta}$ define an open cover of $\mathbb{R}^3$ with the only non-empty intersections being between $\mathbb{R}^3 - (\cup_{p \in \Theta} \bar{B}_{L/4}(p))$ and the balls from $\{B_L(p)\}_{p \in \Theta}$. It follows as a consequence that this collection of sets with the maps $\{g_p\}_{p \in \Theta}$ provide a cocycle definition for a principal SU(2) bundle over $\mathbb{R}^3$. The data $\{a_\infty, \{a_p\}_{p \in \Theta}\}$ supply the corresponding coycle data for a connection on this bundle, this by virtue of (3.8). By the same token, the data $\{\phi_\infty, \{\phi_p\}_{p \in \Theta}\}$ supply the cocyle data for a section of the associated vector bundle with fiber $\mathfrak{su}(2)$. All principal bundles SU(2) bundles on $\mathbb{R}^3$ are isomorphic to the product bundle and so the data $\{(a_\infty, \phi_\infty), \{(a_p, \phi_p)\}_{p \in \Theta}\}$ defines a $C^\infty(\mathbb{R}^3; SU(2))$ equivalence class of connection on the product principal SU(2) bundle over $\mathbb{R}^3$ and section of the product $\mathfrak{su}(2)$ bundle.

*Part 2*: This part of the subsection defines the pair $(a_\infty, \phi_\infty)$. The definition is given momentarily. A digression is needed first to supply needed background. Before starting, introduce by way of notation $\overline{B}_{L/8}(p)$ to denote the concentric, closed radius L/8 ball and use $B_{L/8}(p)$ to denote the open radius L/8 ball centered at p.

The digression begins with the observation that the second integer cohomology of $\mathbb{R}^3 - (\cup_{p \in \Theta} \overline{B}_{L/8}(p))$ is isomorphic to $\mathbb{Z}^N$. More to the point, evaluation on the fundamental class of the boundary of each $p \in \Theta$ version of $\overline{B}_{L/4}(p)$ defines an isomorphism with the group $\oplus_{p \in \Theta} \mathbb{Z}$. Let $\iota_N$ denote the class in the second cohomology that evaluates as 1 on the fundamental class of each $p \in \Theta$ version of $\overline{B}_{L/4}(p)$. The pairing between this class with the fundamental class of any sufficiently large but constant |x| sphere is equal to N.

To continue the digression, let $\mathcal{P}_N \to \mathbb{R}^3 - (\cup_{p \in \Theta} \overline{B}_{L/8}(p))$ denote a principal U(1) bundle whose associated complex line bundle has first Chern class equal to $\iota_N$. By way of a parenthetical remark, any two such bundles are isomorphic because the second cohomology of $\mathbb{R}^3 - (\cup_{p \in \Theta} \overline{B}_{L/8}(p))$ is torsion free. View U(1) as the subgroup of diagonal $2 \times 2$ unitary matrices with determinant 1, and thus as a subgroup of SU(2). Multiplication on the left by this subgroup defines an action of U(1) on SU(2). Use this action to define the principal SU(2) bundle $\mathcal{P}_N \times_{U(1)} SU(2)$ over $\mathbb{R}^3 - (\cup_{p \in \Theta} \overline{B}_{L/8}(p))$. The associated $\mathfrak{su}(2)$ vector bundle is $\mathcal{P}_N \times_{U(1)} \mathfrak{su}(2)$. The latter bundle splits as the direct sum of two bundles, the first being the product line bundle and the second being an oriented 2-plane bundle with Euler class $2\iota_N$. Use $\hat{\sigma}$ to denote the section of $\mathcal{P}_N \times_{U(1)} \mathfrak{su}(2)$ with norm 1 that generates the 1-dimensional real subbundle.

With the digression now over, the definition of $(a_\infty, \phi_\infty)$ starts with the introduction of the harmonic function on $\mathbb{R}^3 - \Theta$ given by the rule

$$x \to 1 - \sum_{p \in \Theta} \frac{1}{|x-p|} .$$

(3.9)

This function is denoted by $\phi_\Theta$. To say that $\phi_\Theta$ is harmonic is to say that $*d\phi_\Theta$ is a closed 2-form. The definition is such that $*d\phi_\Theta$ evaluates as $4\pi$ on the boundary of each $p \in \Theta$ version of $\overline{B}_{L/8}(p)$. It follows as a consequence that $\frac{i}{2} *d\phi_\Theta$ is the curvature 2-form of a connection on $\mathcal{P}_N$. Fix such a connection and use it to induce a connection on the associated principal SU(2) bundle $\mathcal{P}_N \times_{U(1)} SU(2)$. The latter is denoted by $A_\Theta$.

All principal SU(2) bundles over $\mathbb{R}^3 - (\cup_{p \in \Theta} \overline{B}_{L/8}(p))$ are isomorphic to the product bundle. As explained in the next paragraph, an isomorphism from $\mathcal{P}_N \times_{U(1)} SU(2)$ to the product principal SU(2) bundle over $\mathbb{R}^3 - (\cup_{p \in \Theta} \overline{B}_{L/8}(p))$ can be found with the following property: Let $u$ denote the inverse isomorphism. Fix $p \in \Theta$. Then $u^* \hat{\sigma}$ on $B_L(p) - \overline{B}_{L/8}(p)$ is the map to $\mathfrak{su}(2)$ given by the rule

$$x \to \frac{(x-p)_i}{|x-p|} \tfrac{1}{2}\sigma_i.$$

(3.10)

To find an isomorphism with this property, start by choosing some isomorphism from $\mathcal{P}_N \times_{U(1)} SU(2)$ to the product principal $SU(2)$ bundle and denote its inverse by $\mathfrak{h}$. Fix $p \in \Theta$ and let $\mathcal{A}_p$ denote the spherical annulus where $\tfrac{1}{8}L \leq |x-p| \leq L$. Both $\mathfrak{h}^*\hat{\sigma}$ and the map in (3.10) define degree 1 maps from $\overline{B}_{L/4}(p)$ to the unit sphere in $\mathfrak{su}(2)$ and so $\mathfrak{h}^*\hat{\sigma}$ and the map in (3.10) are homotopic maps from $\mathcal{A}_p$ to the unit sphere in $\mathfrak{su}(2)$. It follows as a consequence that there exists a map from $\mathcal{A}_p \to SU(2)$ to be denoted by $h_p$ which is the constant map to the identity where $|x-p| > \tfrac{1}{2}L$ and is such that $h_p(\mathfrak{h}^*\hat{\sigma})h_p^{-1}$ is the map in (3.10) where $|x-p| \leq \tfrac{1}{4}L$. View the collection $\{h_p\}_{p \in \Theta}$ as being a set of automorphisms of the product $SU(2)$ principal bundle over $\mathbb{R}^3 - \cup_{p \in \Theta} \overline{B}_{L/8}(p)$. Keeping in mind that these automorphisms commute, set $u$ to be the composition of first $\mathfrak{h}$ and then $\prod_{p \in \Theta} h_p$.

With $u$ in hand, define $a_\infty$ over $\mathbb{R}^3 - \cup_{p \in \Theta} \overline{B}_{L/8}(p)$ by writing $u^*A_\Theta$ as $\theta_0 + a_\infty$ and define $\phi_\infty$ over this same subset of $\mathbb{R}^3$ to be $\phi_\Theta u^*\hat{\sigma}$.

*Part 3*: This part of the subsection defines each $p \in \Theta$ version of $(a_p, \phi_p)$. The definition uses the number $r_p$ that is defined in (3.6). Keep in mind that $r_p > 0$. The pair $(a_p, \phi_p)$ on the ball $\overline{B}_{\ell/8}(p)$ is given by the formula

$$a_p = \left(\frac{1}{|x-p|} - \frac{r_p}{\sinh(r_p|x-p|)}\right)\varepsilon_{ijk}\frac{(x-p)_i}{|x-p|}dx_j \tfrac{1}{2}\sigma_k \quad \text{and} \quad \phi_p = \left(\frac{r_p}{\tanh(r_p|x-p|)} - \frac{1}{|x-p|}\right)\frac{(x-p)_i}{|x-p|}\tfrac{1}{2}\sigma_i.$$

(3.11)

Note in particular that $(\theta_0 + a_p, \phi_p)$ is the $r = r_p$ version of the pair given in (3.5).

The upcoming definition of $(a_p, \phi_p)$ on the rest of $B_L(p)$ uses notation that is defined directly in this and the subsequent paragraph. The first set of definitions refer to a given point $q \in \Theta - p$. With $q$ in hand, define $\eta_{p,q}$ to be the harmonic function on $B_L(p)$ that is given by the formula

$$x \to \eta_{p,q}(x) = \frac{1}{|x-q|} - \frac{1}{|p-q|} \ .$$

(3.12)

Note in particular that $|\eta_{p,q}| \leq c_0 \frac{|x-p|}{|p-q|^2}$ on $B_L(p)$ because the distance between p and q is no less than 2L. This is also why $|d\eta_{p,q}| \leq c_0 \frac{1}{|p-q|^2}$ on $B_L(p)$. It follows as a consequence of this last bound that the closed 2-form $*d\eta_{p,q}$ can be written on $\overline{B}_{L/4}(p)$ as

$$*d\eta_{p,q} = d\mathfrak{a}_{p,q} \text{ with } \mathfrak{a}_{p,q} \text{ being a 1-form that obeys } |\mathfrak{a}_{p,q}| \leq c_0 \frac{|x-p|}{|p-q|^2} \text{ on } B_L(p) .$$

$$\tag{3.13}$$

The preceding observation has, in turn, the following implication: There is an automorphism of the product principal SU(2) bundle over $B_L(p) - \bar{B}_{L/8}(p)$ with two properties, the first being that it pulls back the connection $\theta_0 + a_\infty$ to

$$\theta_0 + \varepsilon_{ijk} \frac{(x-p)_i}{|x-p|^2} dx_j \tfrac{1}{2}\sigma_k + (\textstyle\sum_{q \in \Theta - p} \mathfrak{a}_{p,q}) \frac{(x-p)_i}{|x-p|} \tfrac{1}{2}\sigma_i \; ;$$

$$\tag{3.14}$$

and the second being that the induced automorphism of the associated $\mathfrak{su}(2)$ bundle has no affect on $\phi_\infty$. Fix such an automorphism and denoted it by $u_p$.

The next definition requires the function $\chi$ from Step 1 in Section 2d. By way of a reminder, this is a chosen non-negative and non-increasing function on $\mathbb{R}$ that equals 1 on $(-\infty, \tfrac{1}{4}]$ and equals zero on $[\tfrac{1}{2}, \infty)$. This function is denoted in what follows by $\chi$. Given $p \in \Theta$, set $\chi_p$ to denote the function $\chi(8L^{-1}|(\cdot) - p| - 1)$ on $\mathbb{R}^3$. This function is equal to zero where $|x-p| > \tfrac{3}{16}L$ and it is equal to one where $|x-p| < \tfrac{1}{8}L$.

The formula that follows defines $(a_p, \phi_p)$ on the whole of $B_L(p)$:

- $a_p = (\frac{1}{|x-p|} - \chi_p \frac{r_p}{\sinh(r_p|x-p|)}) \varepsilon_{ijk} \frac{(x-p)_i}{|x-p|} dx_j \tfrac{1}{2}\sigma_k + (1-\chi_p)(\textstyle\sum_{q \in \Theta - p} \mathfrak{a}_{p,q}) \frac{(x-p)_i}{|x-p|} \tfrac{1}{2}\sigma_i$ .
- $\phi_p = (r_p + \chi_p(\frac{r_p}{\tanh(r_p|x-p|)} - r_p) - \frac{1}{|x-p|} - (1-\chi_p)\textstyle\sum_{q \in \Theta - p} \eta_{p,q} \frac{(x-p)_i}{|x-p|} \tfrac{1}{2}\sigma_i$ .

$$\tag{3.15}$$

It is a consequence of the definitions in the preceding paragraphs that $(a_p, \phi_p)$ as just defined obeys the constraint in (3.8) when $g_p$ is taken to be $u_p$

**c) The choice of the set $\Theta$ for the Bolognese bag solution**

The first three parts of this subsection define the set $\Theta$ and the parameter L that will serve as input for Section 3b's construction of $(A_G, \Phi_G)$. Part 4 discusses the corresponding function $|*F_{A_G} - d_{A_G}\Phi_G|$.

By way of a convention used below and subsequently, what is denoted by $c_0$ signifies a number whose value in any given appearance has no dependence on $\Theta, N, L$, or any other parameters that have been introduced prior to that appearance. The number represented by $c_0$ is than 1 in all cases appearances and it can be assumed to increase between successive appearances.

*Part 1*: The constructions in Section 3b require the choice of a set $\Theta$ that consists of N points in $\mathbb{R}^3$. The version of $\Theta$ used here consists of N points on a sphere about the origin in $\mathbb{R}^3$. To say more about this set, let $R \in [1, \infty)$ denote the radius of this sphere.

Fix a positive integer to be denoted by K. Fix $k \in \{1, \ldots, K\}$. The Euclidean distance between latitude $\frac{k}{K}\pi$ circle and the latitude $\frac{k-1}{K}\pi$ circle is $R\sin\frac{\pi}{2K}$. Keep in mind that this distance can be written as $(1-\mathfrak{e})\frac{R}{2K}\pi$ with $\mathfrak{e}$ being a positive number less than $\frac{\pi^2}{12K^2}$.

For each $k \in \{0, 1, \ldots, K\}$, use $n_k$ to denote the largest of the integers that are less than $2K\sin(\frac{k}{K}\pi)$. The collection of integers $\{n_k\}_{1 \le k \le K-1}$ is such that $n_k = n_{K-k}$ for each k. The smallest of this collection are $n_1$ and $n_{K-1}$, these being equal to 6 when $K \ge c_0$. The numbers in the set $\{n_k\}_{1 \le k \le K/2}$ increase as k from $n_1$ towards $\frac{1}{2}K$.

For each $k \in \{1, 2, \ldots, K-1\}$, fix a set of $n_k$ equally spaced points on the latitude $\frac{k}{K}\pi$ circle by putting the first point on the zero longitude circle and then putting successive points at increasing longitude. Use $\vartheta_k$ to denote the chosen set of $n_k$ points. Note in this regard that the Euclidean distance between nearest neighbor points from $\vartheta_k$ on the latitude $\frac{k}{K}$ circle is $2R\sin(\frac{k}{K}\pi)\sin(\frac{\pi}{n_k})$. Taylor's theorem finds $\sin(\frac{\pi}{n_k}) \ge \frac{9}{10}\frac{\pi}{n_k}$ since $n_k$ is no less than 6. It follows as a consequence that the distance between distinct points from $\Theta_k$ is no smaller than $\frac{R}{2K}\pi$.

The sum $\sum_{1 \le k \le K} n_k$ when $K > c_0$ can be estimated by comparing the sum with the integral from 0 to $\pi$ of the function $\theta \to \frac{2}{\pi}K^2\sin\theta$. The latter integral is $\frac{4}{\pi}K^2$ and it differs from the sum by at most $c_0 K$. It follows as a consequence that $\sum_{1 \le k \le K} n_k$ can be written as $(1+\mathfrak{e}_K)\frac{4}{\pi}K^2$ with $\mathfrak{e}_K$ being a K dependent number with $|\mathfrak{e}_K| \le c_0 K^{-1}$.

The integer N determines the choice of K; it being the smallest positive integer from those where the function $n \to (1+\mathfrak{e}_n)\frac{4}{\pi}n^2$ on $\{1, 2, \ldots\}$ is no less than N. The definition of K implies that

$$K = \tfrac{1}{2}\sqrt{\pi N} + \mathfrak{z}$$

(3.16)

with $|\mathfrak{z}| \le c_0$; and it follows as a consequence that $\sum_{1 \le k \le K} n_k$ is no less than N and no greater than $N + c_0 N^{1/2}$. This last observation has the following implication: The removal of at most $c_0$ points from each set in the collection $\{\vartheta_k\}_{1 \le k \le K}$ gives a new collection of K sets, this denoted by $\{\Theta_k\}_{1 \le k \le K}$, whose union contains precisely N points. If points from $\vartheta_k$ are removed, do this sequentially so that the largest longitude point is removed first, then the point with the second largest longitude, and so on.

*Part 2*: The set $\Theta = \cup_{k \in \{0,1,\ldots K\}}\Theta_k$ can be used for the constructions in Section 4b only if each $p \in \Theta$ version of (3.6)'s number $r_p$ is positive; and such is the case when R is larger than $(1 + c_0 N^{-1/2}\ln N)N$. That this is so follows directly from the upper bound for each $p \in \Theta$ version of $\sum_{q \in \Theta-p}\frac{1}{|p-q|}$ that is supplied by the first bullet of the next lemma.

**Lemma 3.1**: *There exists $\kappa > 1$ with the following significance: Fix $N \geq \kappa$ and $R \geq 1$. Use this data to define the set $\Theta$ as instructed in Part 1 of this subsection. Fix $p \in \Theta$.*

- $\sum_{q \in \Theta - p} \frac{1}{|p-q|} = \frac{N}{R} + \mathfrak{e}_p$ *with $\mathfrak{e}_p$ having norm at most $\kappa \frac{\sqrt{N}}{R} \ln(N)$.*
- $\sum_{q \in \Theta - p} \frac{1}{|p-q|^2} \leq \kappa \frac{N}{R^2} \ln N.$

*Moreover, if $L \geq 1$ and $x$ is any given point in $\mathbb{R}^3$ then*

- $\sum_{q \in \Theta} \frac{1}{|x-q| + L} \leq \frac{N}{R} + \kappa(\frac{1}{L} + \frac{\sqrt{N}}{R} \ln N).$
- $\sum_{q \in \Theta} \frac{1}{(|x-q| + L)^2} \leq \kappa(\frac{1}{L^2} + \frac{1}{N} \ln N).$

Bounds much like those in Lemma 3.1 for similar configurations of N points can be found in [RSZ]. See also [KS] and the references therein.

*Proof of Lemma 3.1*: The proofs of the third and fourth bullets differ only cosmetically from those of the first and second bullets. This being the case, only the proofs of the first and second bullets are given. The four steps that follow prove these first two bullets.

<u>Step 1</u>: Fix $k \in \{0, 1, 2, \ldots\}$ and let p denote a point from $\vartheta$ with latitude $\frac{k}{K} \pi$. A point in $\vartheta$ with latitude from the interval $[\frac{2k-1}{2K} \pi, \frac{2k+1}{2K} \pi]$ must be a point on the $\frac{k}{K} \pi$ circle. Let q denote another point from $\vartheta$ on this circle. The respective longitudinal angles of p and q differ by no less than $2\pi/n_k$. With this understood, let $T_p$ denote the region in the $|x| = R$ sphere where $\theta \in [\frac{2k-1}{2K} \pi, \frac{2k+1}{2K} \pi]$ and where the longitudinal angle differs from p's by $\pi/n_k$. It follows from what was just said that collection $\{T_p\}_{p \in \vartheta}$ are disjoint. It also follows from the definitions that the spherical area of any given $p \in \vartheta$ version of $T_p$ is equal to

$$\frac{2\pi}{K} \sin(\frac{\pi}{2K}) R^2.$$

(3.17)

What with (3.16), this spherical area can be written when $K \geq c_0$ as

$$(1 - \mathfrak{q}) \frac{1}{N} 4\pi R^2,$$

(3.18)

with the absolute value of $\mathfrak{q}$ being less than $c_0 N^{-1/2}$.

There is one further point to be made about the collection $\{T_q\}_{q \in \Theta}$, this being that the complement in the $|x| = R$ sphere of their union is contained in the set of points with distance at most $c_0 N^{1/2}$ from the half of the great circle with zero longitude.

<u>Step 2</u>: This step proves the second bullet of the lemma. To start, note that the Euclidean distance between distinct points from $\Theta$ is no less than $(1 - c_0 N^{-1/2}) (\frac{\pi}{N})^{1/2} R$, this being a consequence of what is said in Step 1. Denote this distance by $\Delta$. Fix $p \in \Theta$ and let $\Theta_\Delta$ denote the set of points in $\Theta$ with distance no less than $2^{10}\Delta$ from $p$.

It follows from the definition of $\Delta$ and what Step 1 said about the set $\{T_q\}_{q \in \vartheta}$ that the sum $\sum_{q \in \Theta_\Delta} \frac{1}{|p-q|^2}$ can be bounded by approximating it as an integral, the result being

$$\sum_{q \in \Theta_\Delta} \frac{1}{|p-q|^2} \leq \frac{N}{R^2} \int_{\Delta/R}^{\pi} (\frac{1}{\sin(\theta/2)})^2 \sin\theta \, d\theta \ . \tag{3.19}$$

More is said about the derivation of (3.19) in Step 3. The integral on the right hand side of (3.19) is equal to $2|\ln(\sin(\Delta/R))|$ as can be seen by writing $\sin\theta$ as $2\sin(\theta/2)\cos(\theta/2)$. This leads to a $c_0 \frac{N}{R^2} \ln N$ bound on the right hand side of (3.19) when $N \geq c_0$. Meanwhile,

$$\sum_{q \in (\Theta - p) - \Theta_\Delta} \frac{1}{|p-q|^2} \leq c_0 \frac{1}{\Delta^2} \tag{3.20}$$

which is less than $c_0 \frac{N}{R^2}$ because $\Delta \geq \frac{R}{\sqrt{N}}$. These bounds give the lemma's first bullet.

<u>Step 3</u>: To explain (3.19), parametrize the radius $R$ sphere centered at the origin by the points in the concentric radius 1 sphere via the map that sends a point $z$ in the latter to the point $Rz$ in the former. Given $q \in \Theta$, let $U_q$ denote the subset of the unit sphere that parametrize the points in $T_q$. It follows from (4.18) that the area of $U_q$ can be written as $(1 - \mathfrak{q}) 4\pi N^{-1}$ with $|\mathfrak{q}| \leq c_0 N^{-1/2}$. The sets $\{U_q\}_{q \in \Theta}$ being pairwise disjoint cover all but a small fraction of the unit sphere, this missing fraction contained in the set of points with distance at most $c_0 N^{-1/2}$ from the half great circle with zero longitude.

To continue, suppose that $q \in \Theta_\Delta$ and that $z \in U_q$. Then Taylor's theorem finds

$$\frac{1}{|p - Rz|} = \frac{1}{|p-q|} + \mathfrak{r} \quad \textit{with} \ |\mathfrak{r}| \leq 2\Delta \frac{1}{|p-q|^2} \ , \tag{3.21}$$

Since $\Delta \leq 2^{-10} |p-q|$ it follows as a consequence that

$$\frac{1}{|p-q|} = (1 + \mathfrak{w}) \frac{1}{4\pi} N \int_{z \in U_q} \frac{1}{|p - Rz|^2} d\Omega \tag{3.22}$$

with the notation such that $d\Omega$ denotes the area 2-form on the unit radius 2-sphere, the integral in question is over the set $U_q$ in this sphere, and $\mathfrak{w}$ is a number with absolute value at most $2^{-5}$ when $N \geq c_0^{-1}$. Let $S_\Delta$ denote the part of the unit radius 2-sphere that parametrizes the points with distance to p greater than $2^{10}\Delta$. Sum the various $q \in \Theta_\Delta$ versions of (3.22) to see that

$$\sum_{q \in \Theta_\Delta} \frac{1}{|p-q|^2} \leq \frac{1}{4\pi}(1 + \frac{1}{32})\frac{N}{R^2} \int_{z \in S_\Delta} \frac{1}{|p/R - z|^2} d\Omega ,$$

(3.23)

this being an inequality by virtue of the fact that $\cup_{q \in \Theta_\Delta} U_q$ is not the whole of $S_\Delta$.

To finish the story, fix spherical coordinates for the unit radius sphere about the origin that has the north pole being p/R. Denote these coordinates by $(\theta, \varphi)$ with $\theta$ being the latitude and $\varphi$ being the longitude. Let z denote a point on the radius 1 sphere about the origin with latitude $\theta$. Then $|p/R - z| = 2\sin(\theta/2)$ and

$$\frac{\pi}{2} \int_{2\sin(\theta/2) > 2^{10}\Delta/R} (\frac{1}{\sin(\theta/2)})^2 \sin\theta \, d\theta = \int_{z \in S_\Delta} \frac{1}{|p/R - z|^2} d\Omega .$$

(3.24)

What is written in (3.19) follows from (3.23) and (3.24) since the domain of integration for the integral in (3.24) is contained in the set where $\theta \geq \Delta/R$ when $N \geq c_0$.

Step 4: This step proves the top bullet of the lemma. Arguments that differ little from those in Steps 2 and 3 can be used to bound $\sum_{q \in \Theta_\Delta} \frac{1}{|p-q|}$ from above and below by approximating it by an integral; the result being

$$\sum_{q \in \Theta_\Delta} \frac{1}{|p-q|} = \frac{N}{2R} \int_0^\pi \frac{1}{2\sin(\theta/2)} \sin\theta \, d\theta + \mathfrak{e}_p$$

(3.25)

with the number $\mathfrak{e}_p$ having three contributions. The first accounts for the error when approximating $\frac{1}{|p/R - z|}$ by $\frac{R}{|p-q|}$ when $Rz \in T_q$. It follows from (3.21) and from the lemma's second bullet that this contribution to $\mathfrak{e}_p$ has absolute value at most $c_0 \frac{\sqrt{N} \ln N}{R}$.

The second contribution to the error is a negative contribution to $\mathfrak{e}_p$ that accounts for the fact that the lower limit of integration in (3.25) is zero rather than being the angle $\theta$ where $2\sin(\theta/2) = 2^{-10}\Delta/R$. This absolute value of this negative contribution to $\mathfrak{e}_p$ is no less than $c_0 \frac{\sqrt{N}}{R}$.

The third contribution to $\mathfrak{e}_p$ is also negative as it accounts for the fact that there are points on the unit 2-sphere that are not on rays through $\cup_{q \in \Theta} T_q$. Let $\mathfrak{u}$ denote this set. As noted previously, $\mathfrak{u}$ is contained in the set of points with distance at most $c_0 N^{-1/2}$ from the half great circle with zero longitude. It follows as a consequence that

$$\int_{z \in U} \frac{1}{|p/R - z|} d\Omega \leq c_0 N^{-1/2},$$

(3.26)

and it follows from (3.26) that the absolute value of this third, negative contribution to $\mathfrak{e}_p$ is no greater than $c_0 \frac{\sqrt{N}}{R}$.

*Part 3*: Fix for the moment $m > 1$ and take

$$R = N(1 + mN^{-1/2} \ln N)$$

(3.27)

The first bullet of Lemma 3.1 has as a corollary that each $p \in \Theta$ version of $r_p$ is positive when $m > c_0$. In fact, if $m > c_0$ and if R is given by (3.27), then it follows from this same bullet that each $p \in \Theta$ version of $r_p$ is such

- $r_p \geq (1 - 2^{-10}) m N^{-1/2} \ln N$.
- $r_p \leq (1 + 2^{-10}) m N^{-1/2} \ln N$.

(3.28)

Fix $m \geq c_0$ so that when R obeys (3.27) then (3.28) holds. Use this same R in Part 2 to define the set $\Theta$. A choice $m \geq c_0 (\ln N)^{c_0}$ is needed for arguments that come later; but assume until told otherwise the smaller lower bound $m > c_0$.

The constructions in Section 3b also require the choice of the number L, this contrained so as to be positive but less than half the separation between any two distinct points from $\Theta$. Part 1 of this section observed that the separation between distinct pairs of points in $\Theta$ is no less than $\frac{R}{2K} \pi$ with K given by (3.16). If $m > c_0$ and if R is given by (3.27), then this minimal separation is larger than $N^{1/2}$. This understood, set $L = m^{-3/4} N^{1/2}$. Note for future reference that

- $L r_p \geq (1 - 2^{-10}) m^{1/4} \ln N$,
- $L r_p \leq (1 + 2^{-10}) m^{1/4} \ln N$,

(3.29)

the being consequence of the bounds in (3.28).

*Part 4*: Use the set $\Theta$ from Part 2 with the parameter L from Part 3 as input for Section 3b's construction of the pair $(A_G, \Phi_G)$. The lemma that follows points out some salient properties of this pair. The lemma uses $\hat{\sigma}$ to denote $|\Phi_G|^{-1} \Phi_G$, this being a norm 1 section of the product $\mathfrak{su}(2)$ bundle that is defined where $\Phi_G \neq 0$.

**Lemma 3.2**: *There exists $\kappa > 1$ with the following significance: Fix $N > \kappa$ and build the pair $(A_G, \Phi_G)$ as instructed in Section 3b using as input $\Theta$ from Part 2 and L from Part 3 as defined using $m > \kappa$. This pair has properties listed below*

- *The zero locus of $\Phi_G$ is the set $\Theta$. Moreover, if x is a point in $\mathbb{R}^3$ with distance L or greater from each point of $\Theta$, then $|\Phi_p|(x) \geq \frac{1}{4} m N^{-1/2} \ln N$. If $p \in \Theta$ and x has distance less than L from p then $|\Phi_G|(x) \geq \kappa^{-1} ( \frac{r_p}{\tanh(r_p|x-p|)} - \frac{1}{|x-p|} )$.*

- *$*F_{A_G} - d_{A_G} \Phi_G$ is zero where the distance to each $p \in \Theta$ is greater than $\frac{1}{4} L$ and where the distance to any given $p \in \Theta$ is less than $\frac{1}{8} L$. Moreover, if $p \in \Theta$, then*
  a) *The norm of $[\hat{\sigma}, (*F_{A_G} - d_{A_G} \Phi_G)]$ is at most $N^{-20}$.*
  b) *The norm of $\langle \hat{\sigma} (*F_{A_G} - d_{A_G} \Phi_G) \rangle$ is at most $\kappa N^{-1} (\ln N)$.*

*Proof of Lemma 3.2*: The proof has four steps.

  Step 1: This step proves the top bullet's assertion about the norm of $\Phi_G$ on $\mathbb{R}^3 - (\cup_{p \in \Theta} B_{L/4}(p))$. The norm here is the absolute value of the function that is depicted in (3.9). The latter is harmonic and thus it takes its minimum value on the boundary of some $p \in \Theta$ version $\overline{B}_{L/4}(p)$. This understood, it is enough to prove that the function depicted in (3.9) is greater than $\frac{1}{2} m N^{-1/2} \ln N$ where the distance to any given $p \in \Theta$ is equal to $\frac{1}{4} L$. To do this, fix $p \in \Theta$ and use what is said in Part 3 of Section 3b about the various $q \in \Theta - p$ versions of (3.12)'s function $\eta_{p,q}$ to conclude that $|\Phi_G|$ on the boundary of $\overline{B}_{L/4}(p)$ is no less than

$$r_p - c_0 L \sum_{q \in \Theta - p} \frac{1}{|p-q|^2} .$$

(3.30)

The expression in (3.30) is no less than $r_p - c_0 m^{-3/4} N^{-1/2} \ln N$ because $L = m^{-3/4} N^{1/2}$ and because of what is said in the second bullet of Lemma 3.1 about the $\Theta - p$ indexed sum that appears in (3.30). Meanwhile, the top bullet in (3.27) asserts that $r_p$ is no less than $(1 - 2^{-10}) m N^{-1/2} \ln N$; and this implies that (3.30)'s expression is no less that $\frac{1}{2} m N^{-1/2} \ln N$ when $m \geq c_0$.

  Step 2: Fix $p \in \Theta$. This step proves the lemma's assertions about $\Phi_G$ on $B_L(p)$. The asserted bound for $|\Phi_G|$ where the distance to p is no greater than $\frac{1}{8} L$ follows immediately from the formula in (3.11). The asserted bound for $|\Phi_G|$ on the $|x-p| \geq \frac{1}{8} L$ part of $B_L(p)$ is obtained with the help of a suitable a priori bound for $\sum_{q \in \Theta - p} |\eta_{p,q}|$. To this end, first use Taylor's theorem to bound each $q \in \Theta - p$ version of $\eta_{p,q}$ on $B_L(p)$ by $c_0 \frac{1}{|p-q|^2} |x - p|$. Use this bound with the second bullet of Lemma 3.1 to see that

$$\sum_{q \in \Theta - p} |\eta_{p,q}| \leq c_0 (N^{-1} \ln N) |x - p|$$

(3.31)

at any given $x \in B_L(p)$.

Keeping the preceding formula in mind, digress for a moment and let x denote a point in $B_L(p)$ with distance at most $c_0^{-1} r_p^{-1}$ from p. Note in this regard that $r_p L$ is much greater than 1 when $N \geq c_0$, this being a consequence of the top bullet in (3.29). In particular, the top bullet in (3.29) has the following implication: If $m > c_0$ and if x is a point with distance between $\frac{1}{8} L$ and L from p, then

$$\frac{r_p}{\tanh(r_p |x-p|)} = r_p + \mathfrak{e}_p$$

(3.32)

with $\mathfrak{e}_p$ having norm bounded $c_0 N^{-100}$. Granted this last observation, a comparison between the lower bound for $r_p$ in the top bullet of (3.28) and the upper bound in (3.31) leads directly to the lemma's claim about $|\Phi_G|$ on the $|x - p| \geq \frac{1}{8} L$ part of $B_L(p)$.

Step 3: This step and the next address the assertions of the second bullet of the lemma. To begin, the fact that $*F_{A_G} - d_{A_G} \Phi_G$ is zero on the complement of $\cup_{p \in \Theta} B_{L/4}(p)$ follows directly from the fact that the function depicted in (3.9) is harmonic. The vanishing of $*F_{A_G} - d_{A_G} \Phi_G$ when the distance to any $p \in \Theta$ is less than $\frac{1}{8} L$ follows because $(a_p, \phi_p)$ on this part of $B_{L/4}(p)$ is a rescaling of the Prasad-Sommerfeld solution.

The proofs of the assertions in Items a) and b) of the lemma's second bullet are given in Step 4. What follows directly sets the stage and notation for the arguments in Step 4. To begin, fix $p \in \Theta$ and define the connection $A_{D,p}$ on the product principal SU(2) bundle over $B_L(p) - p$ to be

$$A_{D,p} = \theta_0 + \varepsilon_{ijk} \frac{(x-p)_i}{|x-p|^2} dx_j \frac{1}{2} \sigma_k$$

(3.33)

this being a translated version of the connection that is depicted in (3.3). The section $\hat{\sigma}$ of the product $\mathfrak{su}(2)$ bundle over $B_L(p) - p$ is $A_{D,p}$ covariantly constant because it is given the rule $x \to \frac{(x-p)_i}{|x-p|} \frac{1}{2} \sigma_i$. Let $\Phi_{D,p}$ denote the section of the product $\mathfrak{su}(2)$ bundle over $B_L(p) - p$ given by the rule

$$\Phi_{D,p} = (r_p - \frac{1}{|x-p|}) \hat{\sigma} ,$$
(3.34)

this being a rescaled and translated version of the section $\Phi_D$ that is depicted in (3.3). the pair $(A_{D,p}, \Phi_{D,p})$ obey (1.1) on $B_L(p)-p$.

A look at (3.15) shows that the connection $\theta_0 + a_p$ can be written over $B_L(p)-p$ as

$$A_{D,p} + \chi_p A_p + \tau_p \mathfrak{a}_p \hat{\sigma} .$$
(3.35)

with $A_p = -\frac{r_p}{\sinh(r_p|x-p|)} \varepsilon_{ijk} \frac{(x-p)_i}{|x-p|} dx_j \frac{1}{2} \sigma_k$, this being an $\mathfrak{su}(2)$-valued 1-form; and with $\mathfrak{a}_p$ denoting the real valued 1-form $\sum_{q \in \Theta-p} \mathfrak{a}_{p,q}$. Note for reference below that $\langle \hat{\sigma} A_p \rangle = 0$. A corresponding depiction of $\phi_p$ writes the latter as

$$\phi_p = \Phi_{D,p} + (\chi_p Q_p + \tau_p \eta_p) \hat{\sigma}$$
(3.36)

with $Q_p = \frac{r_p}{\tanh(r_p|x-p|)} - r_p$ and with $\eta_p = -\sum_{q \in \Theta-p} \eta_{p,q}$.

The $\mathfrak{su}(2)$ valued 1-form $*F_{A_G} - d_{A_G} \Phi_G$ can be written using the notation from (3.35) and (3.36) as a sum of two terms, these being

- $g_T = *(d\chi_p \wedge A_p) + \chi_p(\chi_p - 1)(-Q_p[A_p, \hat{\sigma}]) + *(\mathfrak{a}_p \wedge [\hat{\sigma}, A_p]) + \eta_p[\hat{\sigma}, A_p]))$.
- $g_L = -(*(d\chi_p \wedge \mathfrak{a}_p) - \eta_p d\chi_p) \hat{\sigma} - d\chi_p Q_p \hat{\sigma} + *\chi_p(\chi_p - 1) A_p \wedge A_p$.

(3.37)

Note in particular that $\langle \hat{\sigma} g_T \rangle = 0$ and $[\hat{\sigma}, g_L] = 0$.

<u>Step 4</u>: Consider first Item a) of the second bullet. It follows from what is said subsequent to (3.35) that Item a) concerns only the $g_T$ part of $*F_{A_G} - d_{A_G} \Phi_G$. To bound the norm of $g_T$, look at the formulas for $A_p$ and $Q_p$ to see that their absolute values are no greater than $c_0 r_p e^{-r_p L/c_0}$. It follows from this last bound and (3.29) that $|A_p|$ and $|Q_p|$ are no greater than $c_0 N^{-100}$ if $m$ is greater than $c_0 10^4$. Meanwhile, a look at (3.31) finds $|\eta_p| \leq c_0 m^{-3/4} N^{-1/2} \ln N$ and virtually the same argument that gives (3.31) leads to the bound $|\mathfrak{a}_p| \leq c_0 m^{-3/4} N^{-1/2} \ln N$ also. These bounds imply what is asserted by Item a) of the second bullet.

Item b) of the second bullet follows from a $c_0 N^{-1} \ln N$ bound for $|g_L|$. Since $|d\chi_L|$ is at most $c_0 L^{-1}$, it follows from what was just said about $|\eta_p|$ and $|\mathfrak{a}_p|$ that the norm of the $(*(d\chi_p \wedge \mathfrak{a}_p) - \eta_p d\chi_p) \hat{\sigma}$ part of $g_L$ is at most $c_0 N^{-1} \ln N$. Meanwhile the bounds from the preceding paragraph for the norms of $A_p$ and $Q_p$ lead to a $c_0 N^{-100}$ bound for the norm of what is left of $g_L$.

## 4. Deformations to a true solution

This section constructs a small deformation of Lemma 3.1's pair $(A_G, \Phi_G)$ when $m$ is on the order of $(\ln N)^{20}$ that result in a pair of connection on the product principal SU(2) bundle over $\mathbb{R}^3$ and section of the product $\mathfrak{su}(2)$ bundle that obeys both (1.1) and (1.3) and looks much like a magnetic bag. By way of a look ahead, the construction has much in common with the construction in the author's Ph. D. thesis of solutions to (1.1) and (1.3) for any given N. Even so, substantive differences arise because the norm of $\langle \hat{\sigma}(*F_{A_G} - d_{A_G}\Phi_G) \rangle$ lacks an a priori $\mathcal{O}(N^{-3/2})$ bound.

### a) The deformation equation

To set the stage for what is to come, let $(A, \Phi)$ denote here a given pair of connection on the product principal SU(2) bundle over $\mathbb{R}^3$ and section of the product $\mathfrak{su}(2)$ bundle. Use $g$ to denote the $\mathfrak{su}(2)$ valued 1-form $*F_A - d_A\Phi$.

Use $C$ to denote the space of pairs of the form $(\mathfrak{a}, \eta)$ with $\mathfrak{a}$ being an $\mathfrak{su}(2)$ valued 1-form on $\mathbb{R}^3$ and with $\eta$ being a map to $\mathfrak{su}(2)$. Thus, $C = C^\infty(\mathbb{R}^3; (T^*\mathbb{R}^3 \otimes \mathfrak{su}(2)) \oplus \mathfrak{su}(2))$. By way of an example, the $\mathfrak{su}(2)$ valued 1-form $g$ can be viewed as an element in $C$ with 1-form component $*F_A - d_A\Phi$ and second component equal to zero.

Define the linear, first order differential operator $\mathcal{D}: C \to C$ by the rule whereby a given element $h = (\mathfrak{a}, \eta) \in C$ is sent by $\mathcal{D}$ to the pair whose respective first and second components are

$$*d_A \mathfrak{a} - d_A \eta + [\Phi, \mathfrak{a}] \quad and \quad *d_A * \mathfrak{a} + [\Phi, \eta] .$$

(4.1)

Let $h = (\mathfrak{a}, \eta)$ denote a given element in $C$. The pair $(A', \Phi') = (A + \mathfrak{a}, \Phi + \eta)$ is such that $*F_{A'} - d_{A'} \Phi' = 0$ if the pair $(\mathfrak{a}, \eta)$ obeys an equation that has the schematic form

$$\mathcal{D}h + h\#h + g = 0$$

(4.2)

with $h\#h$ being shorthand for the element in $C$ with $C^\infty(\mathbb{R}^3; T^*\mathbb{R}^3 \otimes \mathfrak{su}(2))$ component given by $*(\mathfrak{a} \wedge \mathfrak{a}) - [\mathfrak{a}, \eta]$ and with zero $C^\infty(\mathbb{R}^3; \mathfrak{su}(2))$ component. If $\lim_{|x| \to \infty} |\Phi| = 1$ then the $|x| \to \infty$ limit of $|\Phi'|$ will be one if $\lim_{|x| \to \infty} |h|$ is zero. If, in addition, $|F_A|^2$ and $|d_A\Phi|^2$ are finite and $(A, \Phi)$ obeys (1.3), then $(A', \Phi')$ will also obey (1.3) if the integrals over $\mathbb{R}^3$ of $|F_{A'}|^2$ and $|d_{A'}\Phi'|^2$ are finite.

Let $\mathcal{D}^\dagger$ denote the formal $L^2$ adjoint of $\mathcal{D}$, this given by replacing $\Phi$ with $-\Phi$ in (4.1). To say what it means to be the formal $L^2$ adjoint, introduce by way of notation $\langle q, q'\rangle$ to denote the pointwise inner product between given elements $q$ and $q'$ in $C$. If $q$ and $q'$ are any two elements with compact support, then the respective integrals over $\mathbb{R}^3$ of $\langle q', \mathcal{D}q\rangle$ and $\langle \mathcal{D}^\dagger q', q\rangle$ are equal.

Theorem 3.1 in Chapter IV of [JT] summaries results from the author's Ph. D. thesis that give sufficient conditions on the pair $(A,\Phi)$ so as to guarantee that (4.1) has a solution that has the form $h = \mathcal{D}^\dagger u$ with $|h|$ being small. A solution of the form $h = \mathcal{D}^\dagger u$ requires that $u$ obey the second order equation

$$\mathcal{D}\mathcal{D}^\dagger u + \mathcal{D}^\dagger u \# \mathcal{D}^\dagger u + g = 0 \ .$$

(4.3)

Given that $|\Phi| \le 1$ as is the case with $\Phi_G$, Theorem 3.1 in Chapter IV of [JT] asserts that (4.3) has a solution if the integrals over $\mathbb{R}^3$ of the $k = \frac{6}{5}$ and $k = 2$ versions $|g|^k$ have a $c_0^{-1}$ bound. The size of these integrals also determine the pointwise norm of $\mathcal{D}^\dagger u$. Unfortunately, the integrals of $|g|^{6/5}$ and $|g|^2$ are likely not small for the $(A_G, \Phi_G)$ version of $g$; the author can say only that these respective integrals are less than $c_0 N^{13/6}$ and $c_0 N^{1/2}$. As explained in the subsequent subsections, the equation in (4.2) can none-the-less be solved with $h$ being small in the cases when $(A, \Phi) = (A_G, \Phi_G)$.

The proof that (4.2) can be solved with $h$ small exploits the Bochner-Weitzenbock formula for the operator $\mathcal{D}\mathcal{D}^\dagger$ given below in (4.4). This formula uses $\nabla_A^\dagger$ to denote the formal, $L^2$ adjoint of $\nabla_A$. The composition of first $\nabla_A$ and then $\nabla_A^\dagger$ gives the covariant Laplace operator $\nabla_A^\dagger \nabla_A$. This Bochner-Weitzenboch formula writes $\mathcal{D}\mathcal{D}^\dagger$ as the sum

$$\mathcal{D}\mathcal{D}^\dagger = \nabla_A^\dagger \nabla_A + [\Phi,[\cdot\,,\Phi]] + \mathfrak{G}(\cdot) \ ,$$

(4.4)

with $\mathfrak{G}$ being an endomorphism that is defined as follows: Write a given $q \in C$ as $(\mathfrak{q},\tau)$ with $\mathfrak{q}$ being the $C^\infty(T^*\mathbb{R}^3 \otimes \mathfrak{su}(2))$ part of $q$ and with $\tau$ denoting the part in $C^\infty(\mathbb{R}^3; \mathfrak{su}(2))$. The corresponding parts of $\mathfrak{G}(q)$ are

$$*(g \wedge \mathfrak{q} + \mathfrak{q} \wedge g) + [g,\tau] \quad and \quad *(*g \wedge \mathfrak{q} - \mathfrak{q} \wedge *g)$$

(4.5)

with $g$ viewed here as an $\mathfrak{su}(2)$ valued 1-form. The Bochner-Weitzenbock formula for $\mathcal{D}^\dagger \mathcal{D}$ is obtained from (4.4) and (4.5) by replacing $g$ by $*F_A + d_A\Phi$.

**b) The linear equation**

The upcoming construction of a solution to (4.3) employs a suitable inverse for the operator $\mathcal{DD}^\dagger$. The upcoming Lemma 4.1 and Proposition 4.2 describes this inverse. To set the notation for Lemma 4.1, write $|\Phi|^{-1}\Phi$ as $\hat{\sigma}$ and use $\|g\|_*$ to denote the norm for $g$ that is defined by the formula

$$\|g\|_* = \sup_{x\in\mathbb{R}^3} |\Phi|^{-2}|\langle \hat{\sigma}\, g\rangle| + \left(\int_{\mathbb{R}^3} |\Phi|^{-3}|[\hat{\sigma}, g]|^3\right)^{1/3}.$$

(4.6)

This definition presupposes that $g$ vanishes near the zero locus of $\Phi$. The proposition also refers to the Banach space completion of the space of compactly supported elements in $C$ using the norm that is denoted by $\|\cdot\|_\mathbb{U}$ and whose square is defined by the rule

$$q \to \|q\|_\mathbb{U}^2 = \sup_{x\in\mathbb{R}^3} |q|^2 + \sup_{x\in\mathbb{R}^3} \int_{\mathbb{R}^3} \frac{1}{|x-(\cdot)|}(|\nabla_A q|^2 + |[\Phi, q]|^2).$$

(4.7)

This Banach space is denoted by $\mathbb{U}$. The proposition refers to a second Banach space, this being the completion of the space of compactly supported elements in $C$ using the norm that is defined by the rule

$$w \to \sup_{x\in\mathbb{R}^3} \int_{\mathbb{R}^3} \frac{1}{|x-(\cdot)|^2} |w|.$$

(4.8)

This last norm is denoted by $\|\cdot\|_\mathbb{W}$ and the associated Banach space is denoted by $\mathbb{W}$.

**Lemma 4.1**: *There exists $\kappa > 1$ with the following significance: Fix a positive integer $N$ and a pair $(A, \Phi)$ of connection on the product principal SU(2) bundle over $\mathbb{R}^3$ and section of the product $\mathfrak{su}(2)$ bundle. Assume that the integrals over $\mathbb{R}^3$ of both $|F_A|^2$ and $|d_A\Phi|^2$ are finite, that $\lim_{|x|\to\infty} |\Phi| = 1$ and that (1.3) is obeyed. Assume that $|\Phi| \leq 1$, that $g$ is zero near $\Phi$'s zero locus and that $\|g\|_* \leq \kappa^{-1}$. Suppose that $w \in C \cap \mathbb{W}$. There exists a unique smooth element $u \in \mathbb{U}$ that obeys the equation $\mathcal{DD}^\dagger u = w$. This solution is such that $\|u\|_\mathbb{U} \leq \kappa^{-1}\|w\|_\mathbb{W}$.*

This lemma is proved momentarily so assume it is true for the moment.

To set the stage for Proposition 4.2, fix $N$ and $m$ so as to invoke Lemma 3.2 and let $(A, \Phi)$ now denote a version of $(A_G, \Phi_G)$ that is described by that lemma. The next paragraph explains why the corresponding version of $g$ has norm $\|g\|_* \le c_0 (\ln N)^{-1}$. Granted that this is so, take $N \ge c_0$ so that Lemma 4.1 can be invoked in the case when $(A, \Phi)$ is Lemma 3.2's pair $(A_G, \Phi_G)$.

To prove that $\|g\|_* \le c_0 (\ln N)^{-1}$, note first that $g$ has support only where the distance to some $p \in \Theta$ is between $\tfrac{1}{8} L$ and $\tfrac{1}{4} L$. It follows from Lemma 3.1's first bullet, the top bullet in (3.28) that $|\Phi_G|$ is greater than $c_0^{-1} m N^{-1/2} \ln N$ on this part of $\mathbb{R}^3$. Meanwhile, Item b) of Lemma 3.1's second bullet asserts the norm of $\langle \hat{\sigma} g \rangle$ is at most $c_0 N^{-1} \ln N$. It follows as a consequence that $|\Phi_G|^{-2} |\langle \hat{\sigma} g \rangle| \le c_0 m^{-1} (\ln N)^{-1}$. Granted what was just said about the norm of $\Phi_G$ on the support of $g$, it follows from Item a) of Lemma 3.2's second bullet that $|\Phi_G|^{-1} |[\hat{\sigma}, g]|$ is at most $c_0 N^{-19}$. Since L is bounded by $c_0 N^{1/2}$, it follows that the volume of the support of $g$ is at most $c_0 N^{5/2}$. The preceding two bounds lead to a $c_0 N^{-18}$ bound for the $L^3$ norm of $|\Phi_G|^{-1} |[\hat{\sigma}, g]|$. Use this bound with the conclusion from the preceding paragraph to bound $\|g\|_*$ by $c_0 (\ln N)^{-1}$.

Suppose that $p \in \Theta$ and that $w_p$ is an element in $C$ that is bounded by a constant multiple of $e^{-r_p |x-p|/2}$. If such is the case, then $w_p$ is in $\mathbb{W}$ and Lemma 4.1 can be invoked obtain an element in $C \cap \mathbb{U}$ to be denoted by $u_p$ that solves the equation $\mathcal{DD}^\dagger u_p = w_p$. The upcoming Proposition 4.2 says more about $u_p$. By way of notation, the proposition uses $r$ to denote the smallest of the numbers $\{r_q\}_{q \in \Theta}$.

**Proposition 4.2**: *The number $\kappa$ in Lemma 3.2 can be chosen so that the following is true: Fix $N > \kappa$ and $m > \kappa (\ln N)^{2/3}$; and let $(A_G, \Phi_G)$ denote the pair from Lemma 3.2. Fix a point $p \in \Theta$. Suppose that $w_p$ is an element in C that vanishes where the distance to $\Theta - p$ is less than $\tfrac{3}{16} L$, and has norm at most 1 where the distance to p is at most $\tfrac{1}{4} L$ and at most $e^{-\tfrac{1}{2} r |x-p|} e^{\tfrac{1}{8} rL}$ at any point $x \in \mathbb{R}^3$ with distance greater than $\tfrac{3}{16} L$ to $\Theta - p$ and distance greater than $\tfrac{1}{4} L$ to p. Granted this assumption about $w_p$, there exists a unique solution in $C \cap \mathbb{U}$ to the equation $\mathcal{DD}^\dagger u = w_p$, this denoted by $u_p$. This solution is such that $h_p = \mathcal{D}^\dagger u_p$ at any given point $x \in \mathbb{R}^3$ obeys*

$$|h_p| \le \kappa m^{-1/8} N^{1/2} (\ln N)^{3/2} \left(\frac{L}{|x-p| + L}\right)^2 .$$

*In addition, if the distance from x to p is greater than $\tfrac{1}{4} L$ and if the distance from x to $\Theta - p$ is greater than $\tfrac{3}{16} L$ from $\Theta - p$, then*

$$|h_{pT}| \le \kappa m^{-5/8} N^{1/2} (\ln N)^{1/2} \left(e^{-\tfrac{1}{2} r|x-p|} e^{\tfrac{1}{8} rL} + \sum_{q \in \Theta - p} \left(\frac{L}{|p-q| + L}\right)^2 e^{-\tfrac{1}{2} r|x-q|} e^{\tfrac{3}{32} rL}\right)$$

The proof of Proposition 4.2 is given in Section 5 of this article. Sections 4d and 4e invoke this proposition to complete the proof of Theorem 1.4.

*Proof of Lemma 4.1*: The proof has four parts.

*Part 1*: This part of the proof serves as a reminder about various dimension 3 Sobolev inequalities. To start, Let $f$ denote an $L^2_1$ function on $\mathbb{R}^3$. The relevant Sobolev inqualities assert that $f$ is an $L^p$ function for p in the interval [2, 6] and that its $L^p$ norm is bounded by a purely p-dependent multiple its $L^2_1$ norm. By way of a reminder, a function $f$ on $\mathbb{R}^3$ is said to be $L^2_1$ when $|f|^2$ and $|df|^2$ have finite integral over $\mathbb{R}^3$. The $L^2_1$ norm of $f$ is the square root of the sum of the latter two integrals. The function $f$ is said to be an $L^p$ function when $|f|^p$ is integrable over $\mathbb{R}^3$. Its $L^p$ norm is the p'th root of the integral of $|f|^p$ over $\mathbb{R}^3$. The $L^p$ norm of a given $L^p$ function $f$ is denoted in what follows by $\|f\|_p$. The second Sobolev inequality asserts the following: Let $f$ denote a function on $\mathbb{R}^3$ with $|df|^2$ being integrable. Then there exists a number $c$ such that $|f-c|$ is in $L^6$ and $\|f-c\|_6$ is bounded by a universal multiple of $\|df\|_2$.

Note that if A is a connection on the product principal SU(2) bundle over $\mathbb{R}^3$ and $\hat{h}$ is an element in $\mathcal{C}$ with $|\nabla_A \hat{h}|$ being square integrable, then $d|\hat{h}|$ is square integrable since the norm of the latter is no greater than $|\nabla_A \hat{h}|$ at any point. It follows as a consequence that there exists $c \in \mathbb{R}$ such that $f = |\hat{h}| - c$ is an $L^6$ function with $L^6$ norm at most $c_0 \|\nabla_A \hat{h}\|_2$.

*Part 2*: Let $\mathbb{H}$ denote the Banach space completion of the space of compactly supported elements in $\mathcal{C}$ using the norm whose square is defined by the rule

$$q \to \|\nabla_A q\|_2^2 + \|[\Phi, q]\|_2^2 \ .$$

(4.9)

Denote this norm by $\|\cdot\|_\mathbb{H}$. The arguments much like those in Chapter IV.4 of [JT] from the author's Ph. D. thesis prove the following: If there exists a number, $c > 0$, such that

$$\|\mathcal{D}^\dagger u\|_2 \geq c^{-1} \|u\|_\mathbb{H},$$

(4.10)

when $u \in C$ has compact support, then the equation $\mathcal{D}\mathcal{D}^\dagger u = w$ has a unique solution in $C \cap \mathbb{H}$ when $w \in C \cap L^{6/5}$. Moreover, this solution has $\|\cdot\|_\mathbb{H}$ norm bounded by $c_0 c \|w\|_{6/5}$. The next paragraphs explain why (4.10) holds with $c = 2$ if $\|g\|_* \leq c_0^{-1}$.

The explanation starts with the introduction of some notation, this being the writing of any given $q \in C$ where $\Phi \neq 0$ as $q = q_T + q_L$ with $\langle \hat{\sigma} q_T \rangle = 0$ and $[\hat{\sigma}, u_L] = 0$. Granted this notation, write $u$ where $\Phi \neq 0$ as $u_T + u_L$ and likewise write $g$ as $g_T + g_L$. It follows from the fact that $\mathfrak{G}$ is defined from commutators with components of $g$ that the point wise inner product between $u$ and $\mathfrak{G}(u)$ is at most

$$c_0 |g_L| |u_T|^2 + c_0 |g_T| |u_T| |u_L|.$$

(4.11)

Use the Bochner-Wietzenboch formula in (4.4) with an integration by parts and (4.11) to see that $\|\mathcal{D}^\dagger u\|_2^2 = \|u\|_\mathbb{H}^2$ when $g$ is everywhere zero and that

$$\|\mathcal{D}^\dagger u\|_2^2 \geq \|u\|_\mathbb{H}^2 - c_0 \int_{\mathbb{R}^3} (|g_L| |u_T|^2 + |g_T| |u_T| |u_L|)$$

(4.12)

when $g$ is not everywhere zero. To make something of (4.12), use the definition of $\|g\|_*$ to bound the integral in (4.12) by

$$c_0 \|g\|_* (\|[\Phi, u_T]\|_2^2 + \|u_L\|_6 \|[\Phi, u_T]\|_2)$$

(4.13)

Keep in mind that $[\Phi, u_T] = [\Phi, u]$ and $\|[\Phi, u]\|_2^2$ is a part of $\|u\|_\mathbb{H}^2$, and that $|u_L| \leq |u|$ and that the $L^6$ norm of $|u|$ is bounded by $c_0 \|u\|_\mathbb{H}$. By way of a reminder this last fact follows from one of Part 1's Sobolev inequalities because $|d|u|| \leq |\nabla_A u|$. In any event, it follows from what was just said about $[\Phi, u_T]$ and about the $L^6$ norm of $|u|$ that the expression in (4.13) is no greater than $c_0 \|g\|_* \|u\|_\mathbb{H}^2$. This last bound with (4.12) lead directly to the lower bound $\|\mathcal{D}^\dagger u\|_2^2 \geq \frac{1}{2} \|u\|_\mathbb{H}^2$ when $\|g\|_* \leq c_0^{-1}$.

*Part 3*: Take $w \in C$ to have compact support. This implies that $w$ is in $L^{6/5}$ and so there exists $u \in C \cap \mathbb{H}$ solving the equation $\mathcal{D}\mathcal{D}^\dagger u = w$. Take the inner product of both sides of the equation $\mathcal{D}\mathcal{D}^\dagger w$ with $\frac{1}{4\pi|x-(\cdot)|} u$. Integrate the resulting identity over $\mathbb{R}^3$. Keeping in mind that $u$ is smooth and an $L^6$ function, integration by parts can be used with the fact that the function $\frac{1}{4\pi|x-(\cdot)|}$ is the Green's function with pole at x for the Laplacian on $\mathbb{R}^3$ to obtain an inequality of the form

$$\tfrac{1}{2}|u|^2(x) + \tfrac{1}{4\pi}\int_{\mathbb{R}^3}\tfrac{1}{|x-(\cdot)|}(|\nabla_A u|^2 + |[\Phi,u]|^2) \leq \tfrac{1}{4\pi}\int_{\mathbb{R}^3}\tfrac{1}{|x-(\cdot)|}|u|\,|w|\ .$$

(4.14)

Since $w$ has compact support, the integral on the right hand side of (4.14) is no greater than the supremum of $|u|$ on the support of $w$ times $\int_{\mathbb{R}^3}\frac{1}{|x-(\cdot)|}|w|$. The fact that $w$ has compact support implies that the latter integral has limit zero as $|x| \to \infty$. This implies that (4.14)'s left hand side defines a bounded function of $|x|$ with upper bound $c_0\|w\|_{\mathbb{W}}^2$.

*Part 4*: Suppose that $w \in C \cap \mathbb{W}$. The exists a sequence $\{w_k\}_{k=1,2,\ldots}$ of compactly supported elements in $C$ that converges to $w$ in $\mathbb{W}$'s Banach space topology and in the $C^1$ topology. It follows as a consequence that there exists a corresponding sequence $\{u_k\}_{k=1,2,\ldots} \subset C \cap \mathbb{H}$ with each $k \in \{1, 2, \ldots\}$ version of $u_k$ obeying $\mathcal{D}\mathcal{D}^\dagger u_k = w_k$. What is said in Part 3 implies that $\|u_i - u_j\|_{\mathbb{U}} \leq c_0 \|w_i - w_j\|_{\mathbb{W}}$ for each pair of indices $i, j \in \{1, 2,\ldots\}$. This implies that $\{u_k\}_{k=1,2,\ldots}$ converges in $\mathbb{U}$ with its limit having norm $\|\cdot\|_{\mathbb{U}}$ at most $c_0\|w\|_{\mathbb{W}}$. Let $u$ denote the limit. Since $\{w_k\}_{k=1,2,\ldots}$ converges to $w$ in $C^1$, ellipitic regularity arguments imply that $\{u_k\}_{k=1,2,\ldots}$ converges in the $C^2$ topology to $u$ and so $u$ is $C^2$. This implies that $u$ obeys the equation $\mathcal{D}\mathcal{D}^\dagger u = w$ pointwise; and since $w$ is smooth, it follows that $u$ is also smooth.

**c) The iteration**

Part 5 of this subsection describes a procedure that constructs a solution to (4.2) with small pointwise norm. The upcoming Proposition 4.3 in Section 4d makes a formal statement to the effect that the construction does indeed yield a suitable solution. The intervening Parts 1-4 of this subsection introduce various notions that needed for the construction and for the statement of Proposition 4.3.

*Part 1*: Fix $p \in \Theta$. A norm on the space of compactly supported elements in $C$ is defined by as follows: Let $q$ denote a given compactly supported element. The norm of $q$ is defined to be the smallest number $z \in [0, \infty)$ such that

- $|q| \leq z(\frac{L}{|x-p|+L})^2$ *for each* $x \in \mathbb{R}^3$.

- $|q_T| \le z(e^{-\frac{1}{2}r|x-p|} e^{\frac{1}{8}rL} + \sum_{q\in\Theta-p}(\frac{L}{|p-q|+L})^2 e^{-\frac{1}{2}r|x-q|} e^{\frac{3}{32}rL})$ *if* $x \in \mathbb{R}^3$ *has distance greater than* $\frac{3}{16}L$ *from* $\Theta-p$ *and distance greater than* $\frac{1}{4}L$ *from* p.

(4.15)

Use $z_p(q)$ to denote the norm of $q$. Let $\mathbb{X}_p$ denote the Banach space closure of the space of compactly supported elements in $C$ using the norm $z_p(\cdot)$. Use $\mathbb{X}$ to denote $\oplus_{p\in\Theta}\mathbb{X}_p$, this being a Banach space with the norm defined by the function $\mathfrak{q} = (q_p)_{p\in\Theta} \to \sup_{p\in\Theta} z_p(q_p)$.

A solution to (4.2) will be constructed that has the form

$$\mathfrak{h} = \sum_{n=0,1,2,\ldots} \sum_{p\in\Theta} \mathfrak{h}_{(n)p}$$

(4.16)

with $\{\mathfrak{q}_{(n)} = (\mathfrak{h}_{(n)p})_{p\in\Theta}\}_{n=0,1,2,\ldots}$ being a sequence in $\mathbb{X}$. This sequence is defined using an inductive algorithm that is described in Part 5. The algorithm requires $m > c_0(\ln N)^{2/3}$ to invoke Proposition 4.2. By way of a look ahead, the corresponding partial sum sequence $\{\sum_{n=0,1,2,\ldots N} \sum_{p\in\Theta} \mathfrak{h}_{(n)p}\}_{N=0,1,\ldots}$ is proved to converge when $m > c_0(\ln N)^{20}$. Parts 2-4 supply some notation and some observations that are needed for what is to come in Part 5.

*Part 2*: Fix $p \in \Theta$ and use $\chi$ to define the function $\tau_{*p}$ to be the function on $\mathbb{R}^3$ given by the rule $x \to \chi(r(|x-p| - \frac{1}{4}L))$. This function is equal to 1 where the distance to p is less than $\frac{1}{4}L$ and it is equal to 0 where the distance to p is greater than $\frac{1}{4}L + r^{-1}$. The set $\{\tau_{*q}\}_{q\in\Theta}$ is used to define the function $\wp_p$ whose value at any given $x \in \mathbb{R}^3$ is

$$\wp_p(x) = \tau_{*p} + \Pi_{q\in\Theta}(1-\tau_{*p}) e^{-\frac{1}{2}r|x-p|} e^{\frac{1}{8}rL} + \sum_{q\in\Theta-p}(\frac{L}{|p-q|+L})^2 (\tau_{*q} + \Pi_{q'\in\Theta}(1-\tau_{*q'})) e^{-\frac{1}{2}r|x-q|} e^{\frac{3}{32}rL}.$$

(4.17)

Supposing that $\mathfrak{h} \in \mathbb{X}_p$, decompose $\mathfrak{h}_T$ as $\sum_{q\in\Theta} \mathfrak{p}_{p,q}(\mathfrak{h}_T)$ with

- $\mathfrak{p}_{p,p}(\mathfrak{h}_T) = (\tau_{*p} + \Pi_{q\in\Theta}(1-\tau_{*q}) e^{-\frac{1}{2}r|(\cdot)-p|} e^{\frac{1}{8}rL}) \frac{1}{\wp_p} \mathfrak{h}_T$.
- $\mathfrak{p}_{p,q}(\mathfrak{h}_T) = (\frac{L}{|p-q|+L})^2 (\tau_{*q} + \Pi_{q'\in\Theta}(1-\tau_{*q'})) e^{-\frac{1}{2}r|(\cdot)-q|} e^{\frac{3}{32}rL}) \frac{1}{\wp_p} \mathfrak{h}_T$ when $q \in \Theta-p$.

(4.18)

These definition are such that any given $q \in \Theta$ version of $\mathfrak{p}_{p,q}(\mathfrak{h}_T)$ is equal to $\mathfrak{h}_T$ where the distance to q is less than $\frac{1}{4}L$, and it is equal to zero where the distance to $\Theta-q$ is less than $\frac{1}{4}L$. In addition,

- $|\mathfrak{p}_{p,p}(\mathfrak{h}_T)| \le c_0 z_p(\mathfrak{h})(\tau_{*p} + \Pi_{q\in\Theta}(1-\tau_{*q}) e^{-\frac{1}{2}r|(\cdot)-p|} e^{\frac{1}{8}rL})$,
- $|\mathfrak{p}_{p,q}(\mathfrak{h}_T)| \le c_0 z_p(\mathfrak{h})(\frac{L}{|p-q|+L})^2 (\tau_{*q} + \Pi_{q'\in\Theta}(1-\tau_{*q'})) e^{-\frac{1}{2}r|(\cdot)-q|} e^{\frac{3}{32}rL}$ when $q \in \Theta-p$;

(4.19)

These bounds follow directly from the definition in (4.15) of $z_p(\cdot)$.

*Part 3*: Suppose that $q = (\mathfrak{a}, \eta)$ and $q' = (\mathfrak{a}', \eta')$ are elements in $C$. Define $q \# q'$ to be the element in $C$ with $C^\infty(\mathbb{R}^3; T^*\mathbb{R}^3 \otimes \mathfrak{su}(2))$ component

$$\tfrac{1}{2} *(\mathfrak{a} \wedge \mathfrak{a}' + \mathfrak{a}' \wedge \mathfrak{a}) - \tfrac{1}{2}([\mathfrak{a}, \eta'] + [\mathfrak{a}', \eta]) ,$$

(4.20)

and with zero $C^\infty(\mathbb{R}^3; \mathfrak{su}(2))$ component. This definition is such that $q \# q'$ is a sum of commutators of components of $q$ with components of $q'$. This being the case, it follows that $(q \# q')_L = q_T \# q_T'$ and $(q \# q')_T = q_T \# q_L' + q_L \# q_T'$.

Suppose now that there exists an element $\mathfrak{h} = (\hat{h}_p)_{p \in \Theta} \subset \mathbb{X}$ such that $q$ can be written as $\sum_{p \in \Theta} \hat{h}_p$. Assume that $q'$ can likewise be written as a sum $q' = \sum_{p \in \Theta} \hat{h}'_p$ with $\mathfrak{h}' = (\hat{h}'_p)_{p \in \Theta}$ being in $\mathbb{X}$. Use this decomposition to write $q \# q'$ as $\sum_{p \in \Theta} x_p$ with each $p \in \Theta$ version of $x_p$ given where $\Phi \neq 0$ by writing $x_p = x_{pL} + x_{pT}$ with

- $x_{pL} = \tfrac{1}{2}(\sum_{q \in \Theta} \mathfrak{p}_{q,p}(\hat{h}_{qT})) \# q'_T + q_T \# (\sum_{q \in \Theta} \mathfrak{p}_{q,p}(\hat{h}'_{qT}))$.
- $x_{pT} = (\sum_{q \in \Theta} \mathfrak{p}_{q,p}(\hat{h}_{qT})) \# q'_L + q_L \# (\sum_{q \in \Theta} \mathfrak{p}_{q,p}(\hat{h}'_{qT}))$.

(4.21)

Note that $x_p$ is a smooth element in $C$ that is equal to zero where the distance to $\Theta - p$ is less than $\tfrac{1}{4} L$. This is so because any given $q \in \Theta - p$ version of $\mathfrak{p}_{q,p}(\hat{h}_{qT})$ is equal to $\hat{h}_{qT}$ on the radius $\tfrac{1}{4} L$ ball about p and it is equal to zero where the distance to $\Theta - p$ is less than $\tfrac{1}{4} L$. A second important obervation follows from (4.19), this being that

- $|x_p| \leq c_0 z(\mathfrak{h}) z(\mathfrak{h}')$ *where the distance to* p *is less than* $\tfrac{1}{4} L$,
- $|x_p| \leq c_0 z(\mathfrak{h}) z(\mathfrak{h}') e^{-\tfrac{1}{2}r|x-p|} e^{\tfrac{1}{8}rL}$ *if* $x \in \mathbb{R}^3$ *is such that* $|x - p| \geq \tfrac{1}{4} L$.

(4.22)

The next part of the subsection proves that these bounds hold.

*Part 4*: To prove the top bullet in (4.22), use the fact that any given $q \in \Theta$ version of $\mathfrak{p}_{q,p}(\hat{h}_{qT})$ is equal to $\hat{h}_{qT}$ on the radius $\tfrac{1}{4} L$ ball centered at p to see that $x_p$ on this ball is equal to $q \# q'$. It follows from the definitions of the norms $\{z_q(\cdot)\}_{q \in \Theta}$ that the norm of any given $q \in \Theta$ version of $\hat{h}_q$ on

$$|\hat{h}_q| \le c_0 z(\mathfrak{h}) \left(\frac{L}{|p-q|+L}\right)^2$$

(4.23)

on the radius $\frac{1}{4}L$ ball centered on p. Granted that this is so, use Lemma 3.1 and the definition of L as $m^{-3/4} N^{1/2}$ to bound $|q|$ by $c_0 z(\mathfrak{h}) m^{-3/2}(\ln N)$ on this ball. Since $m$ is assumed to be greater than $c_0 (\ln N)^{2/3}$, this is at most $c_0 z(\mathfrak{h})$. But for the notation, this argument also proves that $|q'| \le c_0 z(\mathfrak{h}')$ on the radius $\frac{1}{4}L$ ball centered at p. It follows as a consequence that $|q \# q'|$ and thus $|x_p|$ is less than $c_0 z(\mathfrak{h}) z(\mathfrak{h}')$ on this ball.

To prove the second bullet of (4.22), fix $x \in \mathbb{R}^3$ with $|x - p| \ge \frac{1}{4}L$ and the top bullet in (4.15) to see that

$$|q'| \le c_0 z(\mathfrak{h}') \sum_{q \in \Theta} \left(\frac{L}{|x-q|+L}\right)^2 .$$

(4.24)

The fourth bullet of Lemma 3.2 bounds the sum in (4.24) by $c_0 m^{-3/2}(\ln N)$ and this imples a $c_0 z(\mathfrak{h}')$ bound for $|q'|$ because $m \ge c_0 (\ln N)^{2/3}$. Meanwhile, the inequality in (4.19) when applied to the collection $\{\mathfrak{p}_{q,p}\}_{q \in \Theta}$ leads to the bound

$$|\Sigma_{q \in \Theta} \mathfrak{p}_{q,p}(\hat{h}_{qT})| \le c_0 z(\mathfrak{h}) \, e^{-\frac{1}{2}r|x-p|} \, e^{\frac{1}{8}rL} \Sigma_{q \in \Theta} \left(\frac{L}{|p-q|+L}\right)^2 .$$

(4.25)

As just noted, the sum that appears here is at most $c_0 m^{-3/2} \ln N$ and thus at most $c_0$. This bound with the $c_0 z(\mathfrak{h}')$ bound for $|q'|$ bounds the norm of $\Sigma_{q \in \Theta} \mathfrak{p}_{q,p}(\hat{h}_{qT})) \# q'$ by what is written on the right hand side of (4.22). The same argument with the roles of $q$ and $q'$ reversed supplies the same bound for norm of $q \# (\Sigma_{q \in \Theta} \mathfrak{p}_{q,p}(\hat{h}'_{qT}).$

*Part 5*: This part of the subsection describes an inductive algorithm that constructs the desired sequence $\{\mathfrak{q}_{(n)}\}_{n=1,2,\ldots} \subset \mathbb{X}$. This algorithm starts with an initial step to construct $\mathfrak{q}_{(0)}$ and then an induction step that give constructs each $n \ge 1$ version of $\mathfrak{q}_{(n)}$ from $\mathfrak{q}_{(n-1)}$.

INITIAL STEP: This step constructs $\mathfrak{q}_{(0)} = (\hat{h}_{(0)p})_{p \in \Theta}$. To start, use the fact that $g$ in (4.2) is supported where the distance to $\Theta$ is less than $\frac{3}{16}L$ to write $g$ as $\Sigma_{p \in \Theta} g_{(0)p}$ with each $p \in \Theta$ version of $g_{(0)p}$ having support in the radius $\frac{3}{16}L$ ball centered at p. Define $s_p$ to be $\sup_{x \in \mathbb{R}^3} |g_{(0)p}|$. Given $p \in \Theta$, invoke Proposition 4.2 with $w_p = -s_p^{-1} g_{(0)p}$ and let $u_{(0)p}$ denote the corresponding version of Proposition 4.2's element $u_p$. Set $\hat{h}_{(0)p}$ to be $s_p \mathcal{D}^\dagger u_{(0)p}$.

INDUCTION STEP: Suppose that $n \geq 1$ and that $\{q_{(k)}\}_{k=0,1,\ldots,n-1}$ has been constructed. This step constructs $q_{(n)} = (h_{(n)p})_{p\in\Theta}$. To this end, define $h_{(k)}$ for each $k \in \{0, 1, \ldots, n-1\}$ to to be $\sum_{m=0,1,\ldots k}\sum_{p\in\Theta} h_{(k)p}$, this being an element in $C$. Let $g_{(1)}$ denote $h_0 \# h_0$ in the case $n = 1$ and for $n \geq 1$, use $g_{(n)}$ to denote $h_{(n-1)} \# h_{(n-1)} - h_{(n-2)} \# h_{(n-2)}$. A decomposition of $g_{(n)}$ as $\sum_{p\in\Theta} g_{(n)p}$ is given below with each $p \in \Theta$ version of $g_{(n)p}$ having two salient features: Its norm is bounded by a constant multiple of $e^{-\frac{1}{2}r|x-p|} e^{\frac{1}{8}rL}$ at points $x \in \mathbb{R}^3$ with $|x-p| \geq \frac{1}{4}L$ and it vanishes where the distance to $\Theta-q$ is less than $\frac{3}{16}L$. This understood, fix $p \in \Theta$ and a positive number $s_p$ such that $|g_{(n)p}| \leq s_p$ where the distance to $p$ is less than $\frac{1}{4}L$ and such that $|g_{(n)p}| \leq s_p\, e^{-\frac{1}{2}r|x-p|} e^{\frac{1}{8}rL}$ at any point $x \in \mathbb{R}^3$ with $|x-p| \geq \frac{1}{4}L$. Proposition 4.2 can be invoked with $w_p = -s_p^{-1} g_{(n)p}$. Do this and use $u_{(n)p}$ to denote the corresponding version of Proposition 4.2's element $u_p$. Set $h_{(n)p}$ to be $s_p \mathcal{D}^\dagger u_{(n)p}$.

DEFINITION OF $\{g_{(n)p}\}_{p\in\Theta}$: Define $q_{(n-1)} = \sum_{p\in\Theta} h_{(n-1)p}$, this being an element in $C$. If $n = 1$, then $g_{(1)} = h_{(0)} \# h_{(0)}$. Use $\mathfrak{h} = q_{(0)}$ and $\mathfrak{h}' = q_{(0)}$ for the constructions in Part 3 and denote the resulting version of $\{x_p\}_{p\in\Theta}$ by $\{x_{\ddagger p}\}_{p\in\Theta}$. Define $g_{(1)p}$ to be $x_{\ddagger p}$. If $n > 1$, then $g_{(n)}$ can be written as

$$g_{(n)} = 2q_{(n-1)} \# h_{(n-2)} + q_{(n-1)} \# q_{(n-1)}.$$

(4.26)

Define $\mathfrak{h}_{(n-2)}$ to be the element in $\mathbb{X}$ given by $\sum_{k=0,1,\ldots,n-2} q_{(k)}$. Use $\mathfrak{h} = q_{(n-1)}$ and $\mathfrak{h}' = \mathfrak{h}_{n-2}$ for the constructions in Part 3 and use $\{x_{\diamond p}\}_{p\in\Theta}$ to denote the resulting version of $\{x_p\}_{p\in\Theta}$. Take $\mathfrak{h} = q_{n-1}$ and $\mathfrak{h}' = q_{n-1}$ for the constructions in Part 3 and denote the resulting version of $\{x_p\}_{p\in\Theta}$ by $\{x_{\ddagger p}\}_{p\in\Theta}$. Define $g_{(n)p}$ to be $2x_{\diamond p} + x_{\ddagger p}$.

It follows from (4.22) that any given $n \geq 1$ and $p \in \Theta$ version of $|g_{(n)p}|$ is bounded by a constant multiple of $e^{-\frac{1}{2}r|x-p|} e^{\frac{1}{8}rL}$ at $x \in \mathbb{R}^3$ where $|x-p| > \frac{1}{4}L$; and it follows from what is said subsequent to (4.21) that $g_{(n)p} = 0$ where the distance to $\Theta-p$ is less than $\frac{3}{16}L$.

d) **Convergence**

The proposition below states conditions that guarantee the convergence of the partial sum sequence $\{\mathfrak{f}_{(n)} = \sum_{k=0,1,\ldots,n}\sum_{p\in\Theta} h_{(k),p}\}_{n=0,1,\ldots}$ to an element in $C$ that obeys (4.2).

**Proposition 4.3**: *The number $\kappa$ in Lemma 3.2 can be chosen so that the following is true: Fix $N \geq \kappa$ and $m \geq \kappa(\ln N)^{20}$; and let $(A_G, \Phi_G)$ denote the pair from Lemma 3.2. Use the algorithm in Part 5 of Section 4c to define the sequence $\{q_{(n)} = (h_{(n)p})_{p\in\Theta}\}_{n=0,1,\ldots} \subset \mathbb{X}$.*

*The partial sum sequence $\{\mathfrak{h}_{(n)} = \sum_{k=0,1,\ldots,n} \mathfrak{q}_{(n)}\}_{n=0,1,\ldots}$ converges in $\mathbb{X}$ to an element with $z(\cdot)$ norm bounded by $\kappa\, m^{-1/8} (\ln N)^{5/2} N^{-1/2}$. Write this limit as $(\hat{h}_p)_{p \in \Theta}$ and let $\mathfrak{h}$ denote $\sum_{p \in \Theta} \hat{h}_p$. This element $\mathfrak{h}$ is a smooth solution to (4.2).*

**Proof of Proposition 4.3**: The proof uses $c_*$ to denote the larger of the constant $\kappa$ that appears in Proposition 4.2 and the version of $c_0$ that appears in (4.22). There are five parts to the argument.

*Part 1*: Fix $p \in \Theta$. It follows from what is said in Proposition 4.2 that $\hat{h}_{(0)p}$ is in $\mathbb{X}_p \cap C$ and that its $z_p(\cdot)$ norm obeys

$$z_p(\hat{h}_{(0)p}) \leq c_0 c_*\, m^{-1/8} N^{-1/2} (\ln N)^{5/2}.$$
(4.27)

This norm bound follows directly given the $|g_p| \leq c_0 N^{-1} \ln N$ bound from Lemma 3.2. Granted (4.17), write $m$ as $c^8 (\ln N)^{20}$ and assume that $c > c_0$ so that $z_p(\hat{h}_{(0)p}) \leq c^{-1} N^{-1/2}$.

Since the bound $z_p(\hat{h}_{(0)p}) \leq c^{-1} N^{-1/2}$ holds for each $p \in \Theta$, the corresponding element $\mathfrak{q}_{(0)} = (\hat{h}_{(0)p})_{p \in \Theta} \in \mathbb{X}$ has norm $z(\mathfrak{q}_{(0)}) \leq c^{-1} N^{-1/2}$.

*Part 2*: Take $n = 1$ and use (4.22) to see that each $p \in \Theta$ version of $g_{(1)p}$ has norm bounded by $c_* c^{-2} N^{-1}$ at each $x \in \mathbb{R}^3$ and bounded by $c_* c^{-2} N^{-1} e^{-\frac{1}{2} r |x-p|} e^{\frac{1}{8} rL}$ if $|x - p| \geq \frac{1}{4} L$. This being the case, it follows from Proposition 4.2 that the element $\mathfrak{q}_{(1)} = (\hat{h}_{(1)p})_{p \in \Theta}$ in $\mathbb{X}$ has norm $z(\mathfrak{q}_{(1)}) \leq c_*^2 c^{-2} (\ln N)^{-1} N^{-1/2}$. It is a corollary that the norm in $\mathbb{X}$ of $\mathfrak{h}_{(1)} = \mathfrak{q}_{(0)} + \mathfrak{q}_{(1)}$ obeys the bound $z(\mathfrak{h}_{(1)}) \leq (c^{-1} + c^{-2} c_*^2 (\ln N)^{-1}) N^{-1/2}$.

*Part 3*: Suppose that $n \geq 2$ and that any given $k \in \{1, 2, \ldots, n-1\}$ version of $\mathfrak{q}_{(k)}$ has norm in $\mathbb{X}$ that obeys $z(\mathfrak{q}_{(k)}) \leq c^{-k-1} 6^{k-1} c_*^{2k} (\ln N)^{-k} N^{-1/2}$. Granted that this is so, then the partial sum $\mathfrak{h}_{(k)} = \sum_{m=0,1,\ldots,k} \mathfrak{q}_{(k)}$ has norm $z(\mathfrak{h}_{(k)}) \leq c^{-1}(1 + \sum_{m=1,2,\ldots k} c^{-k} 6^{k-1} c_*^{2k} (\ln N)^{-k}) N^{-1/2}$, this being less than $2 c^{-1} N^{-1/2}$ if $N \geq c_0$. It then follows from the definition of $\{g_{(n)p}\}_{p \in \Theta}$ and (4.22) that the norm of each $p \in \Theta$ version of $g_{(n)p}$ is less than $c^{-n-1} 6^{n-1} c_*^{2n-1} (\ln N)^{-n+1} N^{-1}$ at any $x \in \mathbb{R}^3$; and that it is bounded by this same number times $e^{-\frac{1}{2} r |x-p|} e^{\frac{1}{8} rL}$ if $|x - p| \geq \frac{1}{4} L$.

Use these bounds for the norms of the $p \in \Theta$ versions of $\{g_{(n)p}\}$ with what is said in Proposition 4.2 to conclude that the norm of $\mathfrak{q}_{(n)}$ obeys $z(\mathfrak{q}_n) \leq c^{-n-1} 6^{n-1} c_*^{2n} (\ln N)^{-n} N^{-1/2}$.

*Part 4*: If $N \geq c_0$, then the bounds from Part 3 imply that the sequence $\{\mathfrak{h}_{(k)}\}_{k=0,1,\ldots}$ is a Cauchy sequence in $\mathbb{X}$. To see that this is so, let n and m denote given positive integers. Then $z(\mathfrak{h}_{(n+m)} - \mathfrak{h}_{(m)}) \leq \sum_{k=n+1,\ldots,n+m} z(\mathfrak{q}_k) \leq (\sum_{k=n+1,\ldots,n+m} c^{-k-1} 6^{k-1} c_*^{2k} (\ln N)^{-k}) N^{-1/2}$, and thus by $2c^{-n-1} N^{-1/2}$ if $\ln N > 6 c_*^2$. Being a Cauchy sequence, $\{\mathfrak{h}_{(k)}\}_{k=0,1,\ldots}$ has a limit in $\mathbb{X}$ with $z(\cdot)$ norm bounded by $2c^{-1}$. Write this limit as $(\hat{h}_p)_{p\in\Theta}$.

Fix $n \in \{1, 2, \ldots\}$ for the moment. It follows from the bounds in Part 3 that the partial sum $\hat{h}_{(n)} = \sum_{k=0,1,\ldots,n} \sum_{p\in\Theta} \hat{h}_{(k)p}$ obeys an equation of the form

$$\mathcal{D}\hat{h}_{(n)} + \hat{h}_{(n)} \# \hat{h}_{(n)} = e_{(n)},$$

(4.28)

with the norm of $e_{(n)}$ being bounded by $c_0 c^{-n} N^{-1}$ at each $x \in \mathbb{R}^3$ and being bounded by

$$c_0 c^{-n} N^{-1} \sum_{p\in\Theta} e^{-\frac{1}{2}r|x-p|} e^{\frac{1}{8}rL}$$

(4.29)

at any given $x \in \mathbb{R}^3$ with distance greater than $\frac{1}{4}L$ from $\Theta$.

*Part 5*: What is said in (4.28) and (4.29) has as a corollary that $\hat{h} = \sum_{p\in\Theta} \hat{h}_p$ is a weak solution to (4.2) in the following sense: If $t \in C$ has compact support, then

$$\int_{\mathbb{R}^3} (\langle \mathcal{D}^\dagger t, \hat{h} \rangle + \langle t, \hat{h} \# \hat{h} \rangle) = 0$$

(4.30)

Granted (4.30) and that $\hat{h}$ is bounded, then standard elliptic regularity theorems prove that $\hat{h}$ is smooth and thus a smooth solution to (4.2).

### e) Proof of Theorem 1.4

Let $\kappa_*$ denote the version of $\kappa$ that appears in Proposition 4.3. Take $N \geq \kappa_*$. Fix $c > 1$ and take $m = c^8 \kappa_* (\ln N)^{20}$. Invoke Proposition 4.3 using the pair $N$ and $m$; and then write Proposition 4.3's solution $\hat{h}$ to (4.2) as $(\mathfrak{a}, \eta)$ with $\mathfrak{a}$ being the $\mathfrak{su}(2)$ valued 1-form part of $\hat{h}$ and with $\eta$ being the $\mathfrak{su}(2)$ valued function part. The pair $(A_G + \mathfrak{a}, \Phi_G + \eta)$ obeys the left most equation in (1.1) because $\hat{h}$ obeys (4.2). The $z(\cdot)$ norm bound in Proposition 4.3 implies that the norms of $\mathfrak{a}$ and $\eta$ at any given $x \in \mathbb{R}^3$ obey

$$|\mathfrak{a}| + |\eta| \leq c_0 c^{-1} N^{3/2} \frac{1}{|x|^2 + N^2}.$$

(4.31)

The bound for $|\eta|$ in (4.31) implies that $(A_G + \mathfrak{a}, \Phi_G + \eta)$ also obeys the right most equation in (1.1). This being the case, then Part 3 of Section 1 can be invoked to conclude that $|F_{A_G+\mathfrak{a}}|^2$ and thus $|\nabla_{A_G+\mathfrak{a}}(\Phi_G+\eta)|^2$ are integrable on the whole of $\mathbb{R}^3$ and that $(A_G+\mathfrak{a}, \Phi_G+\eta)$ obeys (1.3).

To prove the assertion about $R_\varepsilon$, invoke Lemma 3.2 to see that $|\Phi_G|$ is greater than $\frac{1}{4} m N^{-1/2} \ln N$ where the distance to $\Theta$ is greater than L, this being $m^{-3/4} N^{1/2}$. The bound in (4.31) says that $|\eta| \le c_0 c^{-1} N^{-1/2}$, and so it follows that $|\Phi+\eta|$ is greater than $\frac{1}{8} m N^{-1/2} \ln N$ where the distance to $\Theta$ is greater than L. This implies that the $\varepsilon = \frac{1}{8} m N^{-1/2} \ln N$ version of $R_\varepsilon$ is at most $R + L$, and thus no greater than $N(1 + 2 m N^{-1/2} \ln N)$.

To see about $r_\varepsilon$, note that what is said in (3.28) and Lemma 3.1's third bullet imply that $|\Phi_G| \le c_0 m N^{-1/2} \ln N$ on the $|x| = R$ sphere. Given that $|\eta|$ is at most $c_0 c^{-1} N^{-1/2}$, it follows that $|\Phi_G+\eta| \le c_0 m^{-1/2} \ln N$ on this sphere also. Since $|\Phi_G+\eta|$ obeys (2.6), it can not have a local maximum where $|x| \le R$ and so $|\Phi_G+\eta| < c_0 m^{-1/2} \ln N$ where $|x| \le R$. It follows that the $\varepsilon = c_0 m N^{-1/2} \ln N$ version of $r_\varepsilon$ is at least R, this being $N(1 + m N^{-1/2} \ln N)$.

This same bound for $|\Phi_G+\eta|$ on the $|x| = R$ sphere proves that the $\varepsilon = c_0 m N^{-1/2} \ln N$ version of $\hat{r}_\varepsilon$ is likewise at least $N(1 + m N^{-1/2} \ln N)$.

The assertion about the position of the zeros of $\Phi_G+\eta$ follows from (4.31) and what is said in Lemma 3.2.

## 5. Proof of Proposition 4.2

As noted in the paragraph prior to the statement of the proposition, the element $w_\mathfrak{p}$ is de facto in the space $\mathbb{W}$ because it is bounded at large distances from the origin by the exponential of negative multiple of the distance to the origin. Lemma 4.1 finds a unique solution in $\mathbb{U} \cap C$ to the equation $\mathcal{D}\mathcal{D}^\dagger u = w_\mathfrak{p}$ because $w_\mathfrak{p}$ is in $\mathbb{W} \cap C$. This solution in $\mathbb{W} \cap C$ is $u_\mathfrak{p}$. The subsections that follow derive Proposition 4.2's asserted bounds for $h_\mathfrak{p}$.

### a) Integral and bounds for $h_\mathfrak{p}$

Given $x \in \mathbb{R}^3$, a calculation finds

$$\int_{\mathbb{R}^3} \frac{1}{|x-(\cdot)|} |w_\mathfrak{p}| \le c_0 \frac{L^3}{|x-\mathfrak{p}|+L} \ .$$

(5.1)

Granted (5.1), then what is said by Lemma 4.1 implies that $\|u_\mathfrak{p}\|_\mathbb{U}$ is at most $c_0 L^2$, this being $c_0 m^{-3/2} N$ because $L = m^{-3/4} N$. The bound $\|u_\mathfrak{p}\|_\mathbb{U} \le c_0 m^{-3/2} N$ implies directly that

$$\sup_{x \in \mathbb{R}^3} \int_{\mathbb{R}^3} \frac{1}{|x-(\cdot)|} |h_\mathfrak{p}|^2 \le c_0 m^{-3} N^2$$

(5.2)

because $|\hat{h}|_p$ is no greater than $c_0(|\nabla_A u_p| + |[\Phi, u_p]|)$. The bound in (5.2) implies that

$$\sup_{x \in \mathbb{R}^3} \int_{\mathbb{R}^3} \frac{1}{|x-(\cdot)|^2} |\hat{h}_p|^2 \leq c_0 m^{-2} N^{3/2} \ln N.$$

(5.3)

The four steps that follow derive (5.3).

Step 1: Fix $x \in \mathbb{R}^3$ and let $\iota_x$ denote the function $\chi(r|x-(\cdot)| - 1)$. This function is equal to 1 where the distance to x is less than $r^{-1}$ and equal to zero where the distance to x is greater than $2r^{-1}$. The element $\iota_x \hat{h}_p$ obeys the equation

$$\mathcal{D}(\iota_x \hat{h}_p) = \mathfrak{s}(d\iota_x) \hat{h}_p + \iota_x w_p,$$

(5.4)

with $\mathfrak{s}(\cdot)$ denoting the symbol homomorphism for the operator $\mathcal{D}$. Take the $L^2$ norm of both sides of (5.4); then use the $\mathcal{D}^\dagger \mathcal{D}$ analog of the Bochner-Weitzenboch formula in (4.4) with the bound $\sup_{x \in \mathbb{R}^3} |\nabla_A \Phi| \leq c_0 r^2$ and the bound $|d\iota_x| \leq c_0 r^{-1}$ to see that

$$\|\iota_x \hat{h}_p\|_{\mathbb{H}}^2 \leq c_0 r^2 \|\theta_x \hat{h}_p\|_2^2 + c_0 \|\iota_x w_p\|_2^2,$$

(5.5)

with $\theta_x$ denoting here the characteristic function for the radius $2r$ ball centered at x.

Step 2: Use (5.2) to bound $r^2 \|\theta_x \hat{h}_p\|_2^2$ by $c_0 m^{-3} r N^2$ and use the fact that $w_p$ has has norm at most 1 to bound $\|\iota_x w_p\|_2^2$ by $r^{-3}$. Look at (3.28) to see that the product $rN$ is no greater than $c_0 m N^{-1/2} \ln N$. Use this bound to bound the right hand side of (5.5) and thus $\|\iota_x \hat{h}_p\|_{\mathbb{H}}^2$ by $c_0 m^{-2} N^{3/2} \ln N$.

Step 3: Step 2's bound for $\|\iota_x \hat{h}_p\|_{\mathbb{H}}$ also bounds the $L^2$ norm of $d|\iota_x \hat{h}_p|$ because $|\nabla_A(\cdot)|$ bounds the norm of $d|\cdot|$. Meanwhile, an integration by parts inequality bounds the the $L^2$ norm of the function $\frac{1}{|x-(\cdot)|}|\iota_x \hat{h}_p|$ by twice the $L^2$ norm of $d|\iota_x \hat{h}_p|$. By way of a reminder, this integration by parts inequality asserts the following: Let $f$ denote a continuous function on $\mathbb{R}^3$ with finite $L^6$ norm and with $df$ being square integrable. Then the $L^2$ norm of the product of $f$ and any given $x \in \mathbb{R}^3$ version of the function $\frac{1}{|x-(\cdot)|}$ is no greater than twice the $L^2$ norm of $df$.

Step 4:  Break the integral in (5.3) into two integrals with integrand $\frac{1}{|x-(\cdot)|^2} |\hat{h}_p|^2$, the first with integration domain being the radius $r^{-1}$ ball centered at x and the second with integration domain being the complement of this ball.  It follows from what is said in Steps 2 and 3 that the first of these integrals is no greater than what is written on the right hand side of (5.3).  The second of these integrals is no greater than $r$ times the integral in (5.2) and this is also less than the right hand side of (5.3).

b)  **A sup-norm bound for $\hat{h}_p$**

This part of the proof uses (5.3) to supply an a priori bound for the pointwise norm of $\hat{h}_p$ on the radius $\frac{1}{4}$L ball centered at p, this being the bound

$$|\hat{h}_p| \leq c_0 \, m^{-1/2} \, N^{1/2} \ln N \, .$$
(5.6)

The derivation of this bound has four steps.

Step 1:  Write A on where the distance to p is less than $\frac{1}{2}$L as $\theta_0 + a_p$ with $a_p$ given by the top bullet in (3.15).  Note in particular that $|a_p|$ at x is bounded by $c_0 \frac{r_p}{r_p|x-p|+1}$ and this is bounded by $c_0 r$.  By the same token, the norm of $\Phi$ where the distance to p is less than $\frac{1}{2}$L is also bounded by $c_0 r$.  This is because $\Phi$ is equal to $\phi_p$ with the latter given by the lower bullet of (3.15).

Step 2:  Let $D_0$ denote the operator from $C^\infty(\mathbb{R}^3; T^*\mathbb{R}^3) \oplus C^\infty(\mathbb{R}^3)$ to itself that sends a given pair $(q, \upsilon)$ to $(*dq - d\upsilon, *d*q)$.  Use the connection $\theta_0$ to define $D_0$ as an operator that maps $C$ to itself.  Viewed in this light, the operator $\mathcal{D}$ where the distance to p is less than $\frac{1}{2}$L can be written as $D_0 + \mathfrak{R}_p$ with $\mathfrak{R}_p$ being an endomorphism with norm bounded by $c_0 r$.  This bound on the norm of $\mathfrak{R}_p$ follows directly from the stated bounds on $|a_p|$ and $|\phi_p|$ in Step 1.

Step 3:  Let x denote a point with distance at most $\frac{1}{4}$L from p.  Write $\mathcal{D}$ as in Step 2 so as to write (5.4) as the equation

$$D_0(\iota_x \hat{h}_p) = -\mathfrak{R}_p(\iota_x \hat{h}_p) + \mathfrak{s}(d\iota_x) \hat{h}_p + \iota_x w_p \, .$$
(5.7)

Let y denote a second point in $\mathbb{R}^3$ and let $G_y$ denote the Green's function for the operator $D_0$ with pole at y.  Keep in mind in what follows that $|G_y| \leq c_0 |(\cdot) - y|^{-2}$.  Multiply both sides of (5.7) by $G_y$ and integrate the resulting equality over $\mathbb{R}^3$. Integrate by parts to identify $\iota_x \hat{h}_p$ at y with the respective integrals of the product of $G_y$ with the three terms on the right hand side of (5.7). Denote these integrals by $\mathfrak{e}_\mathfrak{R}$, $\mathfrak{e}_\mathfrak{s}$ and $\mathfrak{e}_w$.

Step 4: The term $\mathfrak{e}_\mathfrak{R}$ is the integral over $\mathbb{R}^3$ of $-G_y \mathfrak{R}_p(\iota_x \hat{h}_p)$. Use the bound on $|G_y|$ and what is said in Section 5a to bound this term by $c_0 r$ times the integral over $\mathbb{R}^3$ of $\frac{1}{|y-(\cdot)|^2} |\iota_x \hat{h}_p|$. Use (5.3) to bound the latter integral by $c_0 r^{-1/2} m^{-1} N^{3/4} (\ln N)^{1/2}$ and so bound $|\mathfrak{e}_\mathfrak{R}|$ by $c_0 r^{1/2} m^{-1} N^{3/4} (\ln N)^{1/2}$. Use (3.28) to bound this in turn by $c_0 m^{-1/2} N^{1/2} \ln N$.

The term $\mathfrak{e}_\mathfrak{s}$ is the integral over $\mathbb{R}^3$ of $G_y \mathfrak{s}(d\iota_x) \hat{h}_p$. Since $|d\iota_x| \leq c_0 r$, a repeat of the argument that was just used to bound $|\mathfrak{e}_\mathfrak{R}|$ also bounds $|\mathfrak{e}_\mathfrak{s}|$ by $c_0 m^{-1/2} N^{1/2} \ln N$.

The term $\mathfrak{e}_w$ is the integral over $\mathbb{R}^3$ of $G_y \iota_x w_p$. Since $w_p$ has norm at most 1 and since the integral is over a ball of radius $2 r^{-1}$, this integral has norm at most $c_0 r^{-1}$, and thus at most $c_0 m^{-1} N^{1/2} (\ln N)^{-1}$.

c) $\{\hat{h}_{pq}\}_{q \in \Theta-p}$ and $\hat{h}_{p\infty}$

Fix $q \in \Theta-p$ and use $\chi$ to construct a nonnegative function that is equal to 1 where the distance to q is less than $\frac{1}{32} L$ and equal to 0 where the distance to q is greater than $\frac{2}{32} L$. This function can and should be constructed so that the norm of its differential is bounded by $c_0 L^{-1}$. Denote this function by $\chi_{*q}$ and set $\hat{h}_{pq} = \chi_{*q} \hat{h}_p$. The latter has compact support in the radius $\frac{1}{16} L$ ball centered at q. Note that $(A_G, \Phi_G)$ where $\chi_{*q}$ is nonzero is the pair that is obtained from what is written in (3.5) by replacing p with q and r with $r_q$. This implies in particular that the $(A_G, \Phi_G)$ version of the operator $\mathcal{D}$ on the support of $\chi_{*q}$ is the identical to that defined by this same version of (3.5).

Use $\chi$ to construct another nonnegative function, this one denoted by $\chi_{\diamond q}$. This function is equal to 1 where the distance to q is less than $\frac{1}{64} L$ and it is equal to 0 where the distance to q is greater than $\frac{1}{32} L$. In addition, the norm of $d\chi_{\diamond q}$ is less than $c_0 L^{-1}$. An important point in what follows is that $d\chi_{\diamond q}$ has support where $\chi_{*q} = 1$ and $d\chi_{*q}$ has support where $\chi_{\diamond q} = 1$.

Use $\chi$ to construct a third nonnegative function, this one denoted by $\tau_p$. The function $\tau_p$ equals 1 where the distance to p is less than $(1 - \frac{1}{100}) \frac{1}{4} L$, it equals 0 where the distance to p is greater than $\frac{1}{4} L$, and $|d\tau_p|$ is bounded by $c_0 L^{-1}$. Introduce by way of notation $\hat{h}_{p\infty}$ to denote $\hat{h}_p - \sum_{q \in \Theta-p} \chi_{\diamond q} \hat{h}_p - \tau_p \hat{h}_p$. This $\hat{h}_{p\infty}$ is zero where the distance to any given $q \in \Theta-p$ is less than $\frac{1}{64} L$ and also where the distance to p is less than $(1 - \frac{1}{100}) \frac{1}{4} L$.

The elements $\hat{h}_{p\infty}$ and $\{\hat{h}_{pq}\}_{q \in \Theta-p}$ obey a coupled system of equations that assert

- $\mathcal{D}\hat{h}_{pq} = \mathfrak{s}(d\chi_{*q})\hat{h}_{p\infty}$ .
- $\mathcal{D}\hat{h}_{p\infty} = -\sum_{q\in\Theta-p}\mathfrak{s}(d\chi_{\diamond q})\hat{h}_{pq} - \mathfrak{s}(d\tau_p)\hat{h}_p + \tau_p w_p$ .

(5.8)

**d) Bounds for $\{\hat{h}_{pq}\}_{q\in\Theta}$**

Fix $q \in \Theta-p$. The three parts of this subsection derive the pointwise bound

$$|\hat{h}_{pq}| \leq c_0 m^{1/2} \ln N \sup_{x\in V_q} |\hat{h}_{p\infty}| .$$

(5.9)

It is likely that $|\hat{h}_{pq}|$ is no greater than $c_0 \sup_{x\in V_q} |\hat{h}_{pq}|$, but the latter bound is not needed for what is to come.

*Part 1*: This part serves as a digression to supply background material for the upcoming derivation of a bound on the norm of the each $q \in \Theta$ version of $\hat{h}_{pq}$. To start the digression, suppose for the moment $p \in \mathbb{R}^3$ is any given point. With $r > 0$, let $(A, \Phi)$ denote the pair that is depicted in (3.5). Let $\mathbb{H}$ denote the corresponding version of the Banach space that is defined in Part 2 of the proof of Lemma 4.1. Invoke what is said in [T] and Chapter IV.4 in [JT] to conclude that the $(A, \Phi)$ version of $\mathcal{D}$ defines a Fredholm operator operator from $\mathbb{H}$ to the Banach space of square integrable elements in $C$ with index equal to 4 and kernel dimension equal to 4. The techniques used in [T] and from the author's Ph. D. thesis (see Chapter IV.10 of [JT]) can also be used to prove that the elements in $\mathbb{H}$ that are annihilated by $\mathcal{D}$ are square integrable. The arguments in [T] and Chapter IV.4 of [JT] that prove $\mathcal{D}$ to be Fredholm imply the following: If $q \in \mathbb{H}$ is $L^2$-orthogonal to the kernel in $\mathbb{H}$ of $\mathcal{D}$, then

$$\|\mathcal{D}q\|_2^2 \geq c_0^{-1} \|q\|_{\mathbb{H}}^2 .$$

(5.10)

A pairwise orthogonal basis for the kernel of $\mathcal{D}$ in $\mathbb{H}$ consists of the element $o_0$ with $C^\infty(\mathbb{R}^3; T^*\mathbb{R}^3 \otimes \mathfrak{su}(2))$ component equal to $\nabla_A \Phi$ and $C^\infty(\mathbb{R}^3; \mathfrak{su}(2))$ component everywhere zero; and then the three elements $\{o_a\}_{a=1,2,3}$ with any given $a \in \{1, 2, 3\}$ version of $o_a$ given by the partial derivative with respect to $x_a$ of the pair that is depicted in (3.11). All four of these elements in $C$ have $L^2$ norm between $c_0^{-1} r^{1/2}$ and $c_0 r^{1/2}$.

Each of the four basis elements can be written as $\mathcal{D}^\dagger u$ with $u \in C$ being an element whose norm has nonzero limit as $|x| \to \infty$. The simplest case is that of $o_0$ as it can be written as $\mathcal{D}^\dagger u_0$ with $u_0$ being the element in $C$ with zero $C^\infty(\mathbb{R}^3; T^*\mathbb{R}^3 \otimes \mathfrak{su}(2))$ component and with $C^{\infty}(\mathbb{R}^3; \mathfrak{su}(2))$ component equal to $-\Phi$. To write $o_a$ as $\mathcal{D}^{\dagger} u_a$, let $a_{p,a}$ denote the $\mathfrak{su}(2)$ valued function given by the $dx^a$ component of (3.11)'s $\mathfrak{su}(2)$ valued 1-form $a_p$. The element $u_a$ has respective $C^{\infty}(\mathbb{R}^3; T^*\mathbb{R}^3 \otimes \mathfrak{su}(2))$ and $C^{\infty}(\mathbb{R}^3; \mathfrak{su}(2))$ components $\Phi \, dx^a$ and $-a_{p,a}$.

*Part 2*: Fix $q \in \Theta - p$ and let $V_q$ denote the spherical shell where the distance to $q$ is greater than $\frac{1}{32}L$ but less than $\frac{2}{32}L$. The top equation in (5.8) implies that

$$\|\mathcal{D} \hat{h}_{pq}\|_2 \leq L^{1/2} \sup_{x \in V_q} |\hat{h}_{p\infty}| \; .$$

(5.11)

With the preceding in mind, define $\{o_a\}_{a=0,1,2,3}$ as in the just concluded Part 1 using the point $q$ in lieu of $p$ and using $r_q$ in lieu of $r$. Let $\hat{h}_{pq\perp}$ denote the $L^2$-orthogonal projection of $\hat{h}_{pq}$ to span of $\{o_a\}_{a=0,1,2,3}$. Use the version of the pair in (3.5) with p replaced by q and with $r$ replaced by $r_q$ to define the Banach space $\mathbb{H}$. The inequality in (5.10) can be invoked with $q$ being $\hat{h}_{pq\perp}$ because the version of $\mathcal{D}$ in the top bullet of (5.8) is identical to the version defined by this same pair. In particular (5.8) and (5.10) lead to the bound

$$\|\hat{h}_{pq\perp}\|_{\mathbb{H}} \leq c_0 L^{1/2} \sup_{x \in V_q} |\hat{h}_{p\infty}| \; .$$

(5.12)

The bound in (5.12) leads in turn to the bound

$$\sup_{x \in \mathbb{R}^3} \left( \int_{\mathbb{R}^3} \frac{1}{|x-(\cdot)|^2} |\hat{h}_{pq\perp}|^2 \right)^{1/2} \leq c_0 L^{1/2} \sup_{x \in V_q} |\hat{h}_{p\infty}| \; .$$

(5.13)

To say more about the derivation of (5.13), note first that the $\|\hat{h}_{pq\perp}\|_{\mathbb{H}}$ bounds the $L^2$ norm of $d|\hat{h}_{pq\perp}|$. The same integration by parts inequality that was invoked in Step 3 in Section 5a bounds the left hand side of (5.13) by twice the $L^2$ norm of $d|\hat{h}_{pq\perp}|$.

As explained directly, (5.12) also leads to a $c_0 L^2 r_q \sup_{x \in V_q} |\hat{h}_{p\infty}|$ bound on the absolute values of the $L^2$ inner products between $\hat{h}_{pq}$ and each $o \in \{o_a\}_{a=0,1,2,3}$. To start the explanation, let $o$ denote the relevant kernel element, and write $o$ as $\mathcal{D}^{\dagger} u$. Having written $o$ in this way, then (5.12) with an integration by parts identifies the $L^2$ inner product between $o$ and $\hat{h}_{pq}$ with the $L^2$ inner product between $u$ and $\mathfrak{s}(d\chi_{*q}) \hat{h}_{p\infty}$. This identification leads directly to the asserted bound for the absolute value of the $L^2$ inner product between any given $o \in \{o_a\}_{a=a=0,1,2,3}$ and $\hat{h}_{pq}$.

*Part 3*: The five steps that follow prove that (5.9) is true.

Step 1: Write A on the support of $h_{pq}$ as $\theta_0 + a_q$ with $a_q$ given by replacing p with q in (3.11). Note in particular that $|a_q|$ at x is bounded by $c_0 \frac{r_q}{r_q|x-q|+1}$ and this is bounded by $c_0 r$. By the same token, the norm of $\Phi$ on the support of $h_{pq}$ is also bounded by $c_0 r$. This is because $\Phi$ on the support of $h_{pq}$ is equal to $\phi_q$ with the latter given by replacing p with q in (3.11).

Step 2: Let $D_0$ again denote the operator from $C^\infty(\mathbb{R}^3; T^*\mathbb{R}^3) \oplus C^\infty(\mathbb{R}^3)$ to itself that sends a given pair $(q, \upsilon)$ to $(*dq - d\upsilon, *d*q)$. As done in Section 5b, use of the connection $\theta_0$ defines $D_0$ as an operator that maps $C$ to itself. With $D_0$ so defined, the operator $\mathcal{D}$ on the support of $h_{pq}$ can be written as $D_0 + \mathfrak{R}_q$ with $\mathfrak{R}_q$ being an endomorphism with norm bounded by $c_0 r$. This bound on the norm of $\mathfrak{R}_q$ follows directly from the stated bounds on $|a_q|$ and $|\phi_q|$ in Step 1.

Step 3: Let x denote a point in the radius $\frac{1}{4}L$ ball centered on q and let $G_x$ denote the Green's function for $D_0$ with pole at x. Multiply both sides of the top bullet in (5.8) with $G_x$ and integrate the resulting equality over $\mathbb{R}^3$. Use the fact that $|G_x| \le c_0|x-(\cdot)|^{-2}$ with Step 2's bound for $|\mathfrak{R}_q|$ to bound the norm $h_{pq}$ at x by a sum of two terms:

$$c_0 r \int_{|(\cdot)-q|<L/32} \frac{1}{|x-(\cdot)|^2} |h_{pq}| \; + \; c_0 L^{-1} \int_{V_q} \frac{1}{|x-(\cdot)|^2} |h_{p\infty}| \; .$$

(5.14)

The right most term in (5.14) is no greater than $c_0 \sup_{x \in V_q} |h_{p\infty}|$. To see about the left most term in (5.14), write $h_{pq}$ as the sum $h_{pq\perp} + \sum_{a=0,1,2,3} z_a o_a$ with $\{z_a\}_{a=0,1,2,3}$ being numbers. The left most term in (5.14) is no greater than a $c_0 r$ times a corresponding sum of five integrals. The first integral in the sum is

$$\int_{|(\cdot)-q|<L/32} \frac{1}{|x-(\cdot)|^2} |h_{pq\perp}| \; ;$$

(5.15)

and the remaining four have the form

$$|z| \int_{\mathbb{R}^3} \frac{1}{|x-(\cdot)|^2} |o| \; ,$$

(5.16)

with o being in turn elements from the set $\{o_a\}_{a=0,1,2,3}$. Meanwhile, z in (5.16) is the corresponding $a \in \{0, 1, 2, 3\}$ version of $z_a$.

Step 4: The bound in (5.13) leads directly to a $c_0 L \sup_{x \in V_q} |\hat{h}_{p\infty}|$ bound for the integral in (5.15). The fact that $|o| \leq c_0 (\frac{r}{r|x-q|+1})^2$ can be used to bound the integral in (5.16) by $c_0 r$. Meanwhile, the absolute value of $z$ is no greater than $c_0 L^2 \sup_{x \in V_q} |\hat{h}_{p\infty}|$. These bounds lead to a $c_0 (rL)^2 \sup_{x \in V_q} |\hat{h}_{p\infty}|$ bound for the left most term in (5.14).

Step 5: The bounds from Steps 3 and 4 lead directly to a $c_0 (1+rL)^2 \sup_{x \in V_q} |\hat{h}_{p\infty}|$ bound for $|\hat{h}_{pq}|$. Since (3.29) says that $rL \leq c_0 m^{1/4} (\ln N)^{1/2}$, this bound implies the asserted $c_0 m^{1/2} \ln N \sup_{x \in V_q} |\hat{h}_{p\infty}|$ bound for $|\hat{h}_{pq}|$.

### e) The definition of $z_L$ and $z_T$

Write $u_{pL}$ as $\hat{u}\hat{\sigma}$ with $\hat{u}$ being a function on $\mathbb{R}^3 - \Theta$. The function $\hat{u}$ obeys the equation $d^\dagger d\hat{u} = \langle \hat{\sigma} w_p \rangle$ where $|x|$ is sufficiently large. Since $u_p$ is in $\mathbb{U}$, its $|x| \to \infty$ limit is zero and so $\hat{u}$ at any point $x \in \mathbb{R}^3$ with $|x|$ sufficiently large is a sum of the integral of $\frac{1}{4\pi|x-(\cdot)|} \langle \hat{\sigma} w_p \rangle$ and a harmonic function with $|x| \to \infty$ limit zero.

Use $c$ to denote a number that can, in principal, depend on p and on N. The value of $c$ can change between successive appearances. Since $|w_p| \leq c\, e^{-r|x|}$ at any point $x \in \mathbb{R}^3$ and since harmonic functions with limit zero have convergent multipole expansions, it follows from what was said in the preceding paragraph that $|\hat{u}| \leq c|x|^{-1}$ and $|d\hat{u}| \leq c|x|^{-2}$ at any given $x \in \mathbb{R}^3$. The key observation in this regard being that

$$\int_{L/4 < |(\cdot)-p|} \frac{1}{|x-(\cdot)|^2} e^{-\frac{1}{2}r|(\cdot)-p|} e^{-\frac{1}{8}rL} \leq c_0 r^{-3} \frac{1}{|x-p|^2} \,.$$

(5.17)

The $c|x|^{-2}$ bound for $|d\hat{u}|$ implies in particular that $|\hat{h}_{p\infty L}|$ at any $x \in \mathbb{R}^3$ is bounded by $c\frac{1}{|x|^2}$ and thus by $(\frac{L}{|x-p|+L})^2$. With the preceding understood, let $z_L$ denote the *smallest* positive number such that the inequality

$$|\hat{h}_{p\infty L}| \leq z_L (\frac{L}{|x-p|+L})^2$$

(5.18)

holds at all points $x \in \mathbb{R}^3$.

As explained momentarily, $|\hat{h}_{pT}|$ is bounded by $ce^{-\frac{1}{2}r|x|}$ at any given point $x \in \mathbb{R}^3$ and thus by $c \sum_{q \in \Theta} e^{-\frac{1}{2}r|x-q|}$. Granted that this is so, let $z_T$ denote the *smallest* positive number such that

- $|\hat{h}_{p\infty T}| \leq z_T(e^{-\frac{1}{2}r|x-p|} e^{\frac{1}{8}rL} + \sum_{q \in \Theta-p}(\frac{L}{|p-q|+L})^2 e^{-\frac{1}{2}r|x-q|} e^{\frac{3}{32}rL})$ *if* $x \in \mathbb{R}^3$ *has distance greater than* $\frac{3}{16}L$ *from* $\Theta-p$ *and* $\frac{1}{4}L$ *from* p.
- $|\hat{h}_{p\infty T}| \leq z_T(\frac{L}{|p-q|+L})^2$ *where the distance to any given* $q \in \Theta-p$ *is less than* $\frac{3}{16}L$ *or where the distance to* p *is less than* $\frac{1}{4}L$.

(5.19)

Keep in mind with regards to (5.19) that the definition of $\hat{h}_{p\infty}$ has it equal to 0 where the distance to $\Theta-p$ is less than $\frac{1}{64}L$ and where the distance to p is less than $\frac{1}{4}L$.

The proof that $|\hat{h}_{pT}| \leq ce^{-\frac{1}{2}r|x|}$ has four steps.

<u>Step 1</u>: The element $u_{p\infty T}$ obeys the equation $\mathcal{D}\mathcal{D}^\dagger u_{pT} = w_{pT}$ where the distance to $\Theta-q$ is greater than $\frac{3}{16}L$ and the distance to p is greater than $\frac{1}{4}L$. The Bochner-Weitzenboch formula in (4.4) for $\mathcal{D}\mathcal{D}^\dagger$ implies that $|u_{pT}|$ on this same domain obeys the differential inequality

$$d^\dagger d |u_{pT}| + r^2 |u_{pT}| \leq |w_{pT}|$$

(5.20)

because $|\Phi|^2 \geq r^2$ where the distance to $\Theta$ is greater than $\frac{3}{16}L$. Note that this bound on $|\Phi|^2$ is a corollary to Lemma 3.2.

<u>Step 2</u>: Suppose that $x \in \mathbb{R}^3$ has distance greater than $3N$ from the origin. Multiply both sides of (5.20) by $\frac{1}{4\pi|x-(\cdot)|}e^{-r|x-(\cdot)|}$ and then integrate the resulting inequality over the complement in $\mathbb{R}^3$ of the radius $2N$ ball centered at the origin. Keeping in mind that $\frac{1}{4\pi|x-(\cdot)|}e^{-r|x-(\cdot)|}$ is the Green's function for the operator $d^\dagger d + r^2$, integration by parts leads to the pointwise inequality

$$|u_{pT}| \leq c_0 \int_{|x|>2N} \frac{1}{|x-(\cdot)|} e^{-r|x-(\cdot)|} |w_{pT}| + ce^{-r|x|} .$$

(5.21)

The right hand side of (5.21) is at most $ce^{-\frac{1}{2}r|x|}$ because $|w_{pT}| \leq c\, e^{-\frac{1}{2}r|(\cdot)|}$. This being the case, $|u_{pT}|$ is also bounded by $ce^{-\frac{1}{2}r|x|}$.

Step 3: Let $\iota_{*x}$ denote the function $\chi(\frac{1}{2} r|x-(\cdot)| - 1)$. This function is equal to 1 where the distance to x is less than $2r$ and it is equal to zero where the distance to x is greater than $4r$. Take the inner product of both sides of the identity $\mathcal{D}\mathcal{D}^\dagger u_p = w_p$ with $\iota_{*x}^2 u_{pT}$ and integrate the resulting identity over the support of $\iota_{*x}$. Use an integration by parts with the pointwise bounds from Step 2 to see that $\|\iota_{*x} u_{pT}\|_\mathbb{H} \le c e^{-\frac{1}{2} r|x|}$. This implies in particular that $c e^{-\frac{1}{2} r|x|}$ bounds the $L^2$ norm of $h_{pT}$ on the radius $2r$ ball centered at x.

Step 4: Repeat Steps 1 and 2 of Section 5a using $h_{pT}$ in lieu of $h_p$. The right hand side of the $h_{pT}$ version of (5.4) is bounded by $c e^{-\frac{1}{2} r|x|}$, this being a consequence of what is said in Step 3 and the pointwise bound for $|w_p|$. It follows that $\|\iota_x h_{pT}\|_\mathbb{H} \le c e^{-\frac{1}{2} r|x|}$. With this understood, the arguments in Steps 3 and 4 of Section 5a can be repeated using $h_{pT}$ in lieu of $h_p$ to prove that $|h_{pT}| \le c e^{-\frac{1}{2} r|x|}$ at points $x \in \mathbb{R}^3$ with $|x|$ being large. Since $|h_{pT}|$ is bounded by $c$ in any event, this implies the claim that $|h_{pT}| \le c e^{-\frac{1}{2} r|x|}$ at any given $x \in \mathbb{R}^3$.

### f) The size of $z_L$

The four parts of this subsection explain why

$$z_L \le \tfrac{1}{50} z_T + c_0 \, m^{-5/8} N^{1/2} (\ln N)^{1/2}$$

(5.22)

when $m > c_0 (\ln N)^{2/3}$.

*Part 1*: To say something about the size of $z_L$, write $h_{p\infty L}$ as $(\mathfrak{a}\hat{\sigma}, \nu\hat{\sigma})$ with $\mathfrak{a}$ being an $\mathbb{R}$-valued 1-form and $\nu$ being an $\mathbb{R}$-valued function. It follows from the second bullet of (4.22) that the latter obey an equation that has the schematic form

$$(*d\mathfrak{a} - d\nu, *d*\mathfrak{a}) = \mathfrak{R}_T(h_{p\infty T}) - \sum_{q\in\Theta-p} \mathfrak{s}(\chi_{\diamond q})\langle\hat{\sigma}\, h_{pq}\rangle - \mathfrak{s}(d\tau_p)\langle\hat{\sigma}\, h_p\rangle + \tau_p \langle\hat{\sigma}\, w_p\rangle$$

(5.23)

with $\mathfrak{R}_T$ being a homomorphism with support where the distance to $\Theta$ is less than $\tfrac{3}{8} L$ and with norm bounded by $c_0 e^{-rL/c_0}$. Note for reference momentarily that the latter bound is less than $N^{-200}$ if $m > c_0$.

Let $D_0$ again denote the operator from $C^\infty(\mathbb{R}^3; T^*\mathbb{R}^3) \oplus C^\infty(\mathbb{R}^3)$ to itself that sends a given pair $(\mathfrak{q}, \upsilon)$ to $(*d\mathfrak{q} - d\upsilon, *d*\mathfrak{q})$. Use the Green's function for $D_0$ to bound $|h_{p\infty L}|$ at any given point x by a sum of three positive terms that are described in the subsequent part of this subsection. These terms are denoted by $\mathfrak{w}_\mathfrak{R}$, $\mathfrak{w}_{\Theta-p}$ and $\mathfrak{w}_p$.

*Part 2*: The $\mathfrak{w}_{\mathfrak{R}}$ term accounts for the appearance of $\mathfrak{R}_T(\hat{h}_{p\infty T})$ on the right hand side of (5.23). It follows from what is said about $\mathfrak{R}_T$ and from the bound in (5.4) that the contribution from $\mathfrak{R}_T(\hat{h}_{p\infty T})$ to the norm of $\hat{h}_{p\infty L}$ at x is no greater than

$$c_0 \sum_{q \in \Theta} \left( \frac{L}{|x-q| + L} \right)^2 N^{-190} .$$
(5.24)

Note that this is at most $c_0 N^{-100} \left( \frac{L}{|x-p| + L} \right)^2$ because p has distance at most 2N from any given element in $\Theta$.

The term $\mathfrak{w}_{\Theta-p}$ is a sum of terms that are indexed by the points in $\Theta-p$ with the term indexed by any given point q accounting for the appearance of $\mathfrak{s}(d\chi_{\diamond q})\langle \hat{\sigma}\, \hat{h}_{pq} \rangle$ on the right hand side of (5.23). The contribution to $\mathfrak{w}_{\Theta-p}$ from q's term is no greater than

$$c_0 L^{-1} \int_{L/64 < |(\cdot)-q| < L/32} \frac{1}{|x-(\cdot)|^2} |\hat{h}_{pq}| .$$
(5.25)

The integral in (5.25) is bounded by writing $\hat{h}_{pq}$ as the sum of $\hat{h}_{pq\perp}$ and its $L^2$-orthogonal projection to the span of $\{o_a\}_{a=0,1,2,3}$. The contribution to (5.25) from $\hat{h}_{pq\perp}$ is at most

$$c_0 (z_L + z_T) \left( \frac{L}{|x-q| + L} \right)^2 \left( \frac{L}{|p-q| + L} \right)^2 ,$$
(5.26)

this being a consequence of (5.13) with the definitions in (5.18) and (5.19) of $z_L$ and $z_T$.

To see about the contribution from the projection of $\hat{h}_{pq}$ to the span of $\{o_a\}_{a=0,1,2,3}$, let $o$ denote any one of these four elements. As noted in Part 1 of Section 5d, the $L^2$ norm of $o$ is between $c_0^{-1} r_q^{1/2}$ and $c_0 r_q^{-1/2}$. Given this fact and given what is said in the final paragraph of Part 2 in Section 5d, it follows that the $L^2$-orthogonal projection of $\hat{h}_{pq}$ to the span of $o$ can be written as $z_o\, o$ with $z_o$ bounded by $c_0 L^2 \sup_{x \in V_q} |\hat{h}_p|$. As noted previously, $\hat{h}_p = \hat{h}_{p\infty}$ on $V_q$, and it follows as a consequence that $z_o$ is at most $c_0 L^2 (z_T + z_L)\left( \frac{L}{|p-q| + L} \right)^2$. This bound with the fact that the norm of $o$ on the domain of the integral in (5.24) is bounded by $c_0 L^{-2}$ implies that $o$'s contribution to (5.25) is also bounded by the expression in (5.26).

The third term, $\mathfrak{w}_p$, accounts for the appearance of the terms $\mathfrak{s}(d\tau_p)\langle \hat{\sigma}\, \hat{h}_{pL} \rangle$ and $\tau_p \langle \hat{\sigma}\, w_p \rangle$ on the right hand side of (5.21). This term is bounded by

$$c_0 L^{-1} \int_{(1-t)L/4 < |(\cdot)-p| < L/4} \frac{1}{|x-(\cdot)|^2} |\hat{h}_p| \;+\; c_0 \int_{(1-t)L/4 < |(\cdot)-p|} \frac{1}{|x-(\cdot)|^2} |w_p| .$$

(5.27)

The bound in (5.2) when $|x-p| > L$ and that (5.3) when $|x-p| \le L$ can be used to bound the left most term in (5.27) by

$$c_0 \left(\frac{L}{|x-p|+L}\right)^2 m^{-5/8} N^{1/2} (\ln N)^{1/2} .$$

(5.28)

Since $|w_p| \le 1$ where the distance to p is less than $\tfrac{1}{4} L$ and since $L = m^{-3/4} N^{1/2}$, this also bounds the contribution to the right most term in (5.27) from the ball of radius $\tfrac{1}{4} L$ centered at p. Meanwhile, (5.17) with the fact that $L \ge r^{-1}$ implies that (5.28) bounds the contribution to the right most integral in (5.27) from the complement of the radius $\tfrac{1}{4} L$ ball centered at p.

*Part 3*: The preceding bounds for $w_{\mathfrak{R}}$, $w_{\Theta-p}$ and $w_p$ leads directly to the bound

$$|\hat{h}_{p\infty L}| \le c_0 (z_L + z_T) \sum_{q \in \Theta-p} \left(\frac{L}{|x-q|+L}\right)^2 \left(\frac{L}{|p-q|+L}\right)^2 + c_0 \left(\frac{L}{|x-p|+L}\right)^2 m^{-5/8} N^{1/2} (\ln N)^{1/2}$$

(5.29)

when $m > c_0$. As explained momentarily, the $\Theta-p$ indexed sum in (5.29) is at most

$$c_0 (z_L + z_T) \, m^{-3/2} \ln N \left(\frac{L}{|x-p|+L}\right)^2 .$$

(5.30)

If $m \ge c_0 (\ln N)^{2/3}$, then this is at most $\tfrac{1}{100} (z_L + z_T) \left(\frac{L}{|x-p|+L}\right)^2$. Granted this lower bound for $m$, then the inequality in (5.30) has the corollary that

$$|\hat{h}_{p\infty L}| \le \left(\tfrac{1}{100} (z_L + z_T) + c_0 \, m^{-5/8} N^{1/2} (\ln N)^{1/2}\right) \left(\frac{L}{|x-p|+L}\right)^2 .$$

(5.31)

This inequality leads directly to the bound in (5.22) for $z_L$ because (5.22) follows from any version of (5.31) that uses for x a point where $|\hat{h}_{p\infty L}|$ is greater than $\tfrac{9}{10} z_L \left(\frac{L}{|x-p|+L}\right)^2$.

*Part 4*: To see why (5.30) bounds the $\Theta-p$ indexed sum on the right hand side of (5.29), separate the sum into two sums. The first sum is indexed by the points $q \in \Theta-p$ with the property that $|x-q| \ge \tfrac{1}{2} |x-p|$; and the second sum is indexed by the points where this inequality fails. The appearances of $(|x-q|+L)^{-2}$ in the first sum can be replaced by $c_0 (|x-p|+L)^{-2}$ with the result being the larger sum

$$\left(\frac{L}{|x-p| + L}\right)^2 \sum_{q \in \Theta} \left(\frac{L}{|p-q| + L}\right)^2.$$

(5.32)

The sum in (5.32) is bounded by $c_0 L^2 N^{-1} \ln N$, this a consequence of what is said by the fourth bullet of Lemma 3.1. It follows as a consequence that the first sum is bounded by what is written in (5.30).

The second sum contains only points $q \in \Theta - p$ with $|x - q| \leq \frac{1}{2}|x - p|$. The latter inequality requires that $|p - q| \geq |x - p|$. This being the case the second sum is at most

$$\left(\frac{L}{|x-p| + L}\right)^2 \sum_{q \in \Theta} \left(\frac{L}{|x-q| + L}\right)^2.$$

(5.33)

As with the first sum, the assertion in the fourth bullet of Lemma 3.1 implies that what is written in (5.33) is no greater than what is written in (5.30).

### g) The size of $h_{p\infty T}$

The six parts of this subsection proves that $z_T$ obeys the bound

$$z_T \leq \tfrac{1}{50} z_L + c_0 m^{-5/8} N^{1/2} (\ln N)^{1/2}.$$

(5.34)

when $m \geq c_0 (\ln N)^{2/3}$. This inequality with the one in (5.22) require that $z_L$ and $z_T$ be less than $c_0 m^{-5/8} N^{1/2} (\ln N)^{1/2}$. This conclusion with the definitions in (5.18) and (5.19) of $z_L$ and $z_T$, the bound in (5.29) asserting $|\hat{h}_{pq}| \leq c_0 m^{1/2} \ln N \sup_{x \in V_q} |\hat{h}_{p\infty}|$, and the bound in (5.6) asserting $|\hat{h}_p| \leq c_0 m^{-1/2} N^{1/2} \ln N$ imply what is asserted by Proposition 4.2.

*Part 1*: A pair of connection on the product SU(2) bundle over $\mathbb{R}^3 - \Theta$ and section of the product $\mathfrak{su}(2)$ bundle over this same domain is defined by the gluing data that uses the same pair $(a_\infty, \phi_\infty)$ from Part 2 of Section 3b but uses for each $p \in \Theta$ the pair $(a_p, \phi_p)$ that is defined by taking $\chi_p = 0$ version of (3.15). Denote this pair by $(A_\infty, \Phi_\infty)$. Note in particular that the section $\hat{\sigma}$ is $A_\infty$-covariantly constant. The pair $(A, \Phi)$ on the support of $h_{p\infty}$ can be written as $(A_\infty, \Phi_\infty) + \sum_{q \in \Theta} b_q$ with each $q \in \Theta$ version of $b_q$ having support where the distance to $q$ is less than $\tfrac{3}{16} L$. Moreover, the norm of $b_q$ is bounded by $c_0 N^{-200}$ on the support of $h_{p\infty}$ when $m > c_0$. The notation in what follows uses $\mathcal{D}_\infty$ to denote the $(A_\infty, \Phi_\infty)$ version of the operator $\mathcal{D}$.

Use $\mathcal{D}_\infty$ to write the $\hat{\sigma}$-orthogonal part of the bottom equation in (5.8) as

$$\mathcal{D}_\infty h_{p\infty T} = \sum_{q \in \Theta} (\mathfrak{F}_{qL}(h_{p\infty T}) + \mathfrak{F}_{qT}(h_{p\infty L})) - \sum_{q \in \Theta - p} \mathfrak{s}(d\chi_{\diamond q}) h_{pqT} - \mathfrak{s}(d\tau_p) h_{pT} + \tau_p w_{pT}$$

(5.35)

with any given $q \in \Theta$ version of $\mathfrak{F}_{qL}$ and $\mathfrak{F}_{qT}$ being endomorphisms with norms bounded by $c_0 N^{-200}$ and with support in the radius $\frac{3}{16}L$ ball centered at q. Act on both sides of this equation by $\mathcal{D}_\infty^\dagger$ and use the Bochner-Weitzenboch formula for $\mathcal{D}_\infty^\dagger \mathcal{D}_\infty$ to obtain a second order equation that can be written schematically as

$$\nabla_{A_\infty}^\dagger \nabla_{A_\infty} \hat{h}_{p\infty T} + |\Phi_\infty|^2 \hat{h}_{p\infty T} + \mathfrak{G}_{+\infty}(\hat{h}_{p\infty T}) = \mathcal{D}^\dagger \hat{k}$$

(5.36)

with $\hat{k}$ denoting the right hand side of (5.35) and with $\mathfrak{G}_{+\infty}$ defined by taking $g$ in (4.5) equal to $2\nabla_{A_\infty}\Phi$. Note that Lemma 3.2 implies that any $q \in C$ version of $\mathfrak{G}_{+\infty}(q)$ has norm at most $\frac{1}{100}|\Phi_\infty|^2|q|$ on the support of $\hat{h}_{p\infty}$ when $m > c_0$ and $N > c_0$. It is also the case that $|\Phi_\infty| \geq r$ on the support of $\hat{h}_{p\infty T}$, this being another consequence of Lemma 3.2. These bounds for $\mathfrak{G}_{+\infty}(\cdot)$ and $|\Phi_\infty|$ are assumed in what follows.

*Part 2*: Take the inner product of both sides of (5.36) with $\hat{h}_{p\infty T}$ and use what was just said about the size of $\mathfrak{G}_{+\infty}$ to obtain the differential inequality

$$\tfrac{1}{2} d^\dagger d |\hat{h}_{p\infty T}|^2 + \tfrac{3}{4} r^2 |\hat{h}_{p\infty T}|^2 \leq \langle \hat{h}_{p\infty T}, \mathcal{D}_\infty^\dagger \hat{k} \rangle$$

(5.37)

with $\langle \hat{h}_{p\infty T}, \mathcal{D}^\dagger \hat{k} \rangle$ denoting the inner product between $\hat{h}_{p\infty T}$ and $\mathcal{D}_\infty^\dagger \hat{k}$.

Fix $x \in \mathbb{R}^3$, multiply both sides of (5.38) by $\frac{1}{4\pi|x-(\cdot)|}e^{-r|x-(\cdot)|}$ and then integrate the resulting inequality over $\mathbb{R}^3$. Since $\frac{1}{4\pi|x-(\cdot)|}e^{-r|x-(\cdot)|}$ is the Green's function with pole at x for the operator $d^\dagger d + r^2$ with pole at x, integration by parts leads to an inequality for $|\hat{h}_{p\infty T}|^2$ at x that reads

$$|\hat{h}_{p\infty T}|^2 + \tfrac{1}{4}r^2 \int_{\mathbb{R}^3} \tfrac{1}{|x-(\cdot)|} e^{-\sqrt{\tfrac{5}{4}}r|x-(\cdot)|} |\hat{h}_{p\infty T}|^2 \leq c_0 \int_{\mathbb{R}^3} \tfrac{1}{|x-(\cdot)|} e^{-\sqrt{\tfrac{5}{4}}r|x-(\cdot)|} |\hat{k}|^2 +$$

$$c_0 \int_{\mathbb{R}^3} \tfrac{1}{|x-(\cdot)|} e^{-\sqrt{\tfrac{5}{4}}r|x-(\cdot)|} \left(\tfrac{1}{|x-(\cdot)|} + r\right) |\hat{h}_{p\infty T}||\hat{k}| \ .$$

(5.38)

The appearance of the $r^2$ term on the *left hand* side of (5.38) can be used to replace the term with the factor $r|\hat{h}_{p\infty T}||\hat{k}|$ in the right most integral on the *right hand* side of (5.38) with $c_0$ times the left most integral on the right hand side of (5.38). Make this replacement to obtain the somewhat simpler looking inequality

$$|\hat{h}_{p\infty T}|^2 \leq c_0 \int_{\mathbb{R}^3} \frac{1}{|x-(\cdot)|} e^{-\sqrt{\frac{5}{4}}r|x-(\cdot)|} |k|^2 + c_0 \int_{\mathbb{R}^3} \frac{1}{|x-(\cdot)|^2} e^{-\sqrt{\frac{5}{4}}r|x-(\cdot)|} |\hat{h}_{p\infty T}||k| \ .$$

(5.39)

Both of the integrals on the right hand side can be written as a sum of integrals that are indexed by the points in $\Theta$. The integrands for the integrals in these sums that are indexed by any given $q \in \Theta - p$ have support in the radius $\frac{3}{16}L$ about the q. The integrands for the integrals in these sums that are indexed by p have support where the distance to $\Theta - p$ is greater than $\frac{3}{16}L$. Note in this regard that the $k = w_p$ where the distance to $\Theta - p$ is greater than $\frac{3}{16}L$ and the distance to p is greater than $\frac{1}{4}L$.

*Part 3*: Suppose that $q \in \Theta - p$. The contribution to the left most integral on the right hand side of (5.39) from the radius $\frac{3}{16}L$ ball centered on q is no greater than

$$\int_{|(\cdot)-q|<3L/16} \frac{1}{|x-(\cdot)|} e^{-r|x-(\cdot)|} (|\mathfrak{F}_{qL}|^2 |\hat{h}_{p\infty T}|^2 + |\mathfrak{F}_{qT}|^2 |\hat{h}_{p\infty L}|^2 + |d\chi_{\diamond q}|^2 |\hat{h}_{pq}|^2) \ .$$

(5.40)

The subsequent analysis exploits the fact that $|\mathfrak{F}_{qL}| \leq N^{-200}$ and $|\mathfrak{F}_{qT}| \leq N^{-200}$, and the fact that the support of $d\chi_{\diamond q}$ is in the radius $\frac{1}{32}L$ ball centered at q. Granted these facts, it then follows directly from the $|\hat{h}_{pq}|$ bound in (5.9) and the definitions of $z_L$ and $z_T$ in (5.18) and (5.19) that the $|x-q| > \frac{3}{16}L$ version of (5.40) is no greater than

$$c_0 N^{-100} (z_T^2 + z_L^2) e^{-r|x-q|} e^{\frac{3}{16}rL} \left(\frac{L}{|p-q|+L}\right)^4$$

(5.41)

in the case when $m > c_0$.

Now suppose that $|x-q| \leq \frac{3}{16}L$. The just stated bounds $|\mathfrak{F}_{qL}| \leq N^{-200}$ and $|\mathfrak{F}_{qT}| \leq N^{-200}$ with the definitions in (5.18) and (5.19) lead to bounds on the respective contributions to (5.40) from the sum $|\mathfrak{F}_{qL}|^2 |\hat{h}_{p\infty T}|^2 + |\mathfrak{F}_{qT}|^2 |\hat{h}_{p\infty L}|^2$ by

$$c_0 N^{-200} (z_T^2 + z_L^2) \left(\frac{L}{|p-q|+L}\right)^4 \ .$$

(5.42)

Write $\hat{h}_{pq}$ as $\hat{h}_{pq\perp} + \sum_{a=0,1,2,3} z_a o_a$ to obtain a suitable bound for the $|d\chi_{\diamond q}|^2 |\hat{h}_{pq}|^2$ contribution to (5.40) in the case when $|x-p| \leq \frac{3}{16}L$. By way of a reminder, $\{z_a\}_{a=0,1,2,3}$ is a set of numbers, $\{o_a\}_{a=0,1,2,3}$ is a set in $C \cap L^2$ that is defined in Part 1 of Section 5d and $\hat{h}_{pq\perp}$ is the $L^2$ projection of $\hat{h}_{pq}$ to the $L^2$ orthogonal complement of the span of $\{o_a\}_{a=0,1,2,3}$. Use this depiction of $\hat{h}_{pq}$ to see that the $|d\chi_{\diamond q}|^2 |\hat{h}_{pq}|^2$ contribution to (5.40) is at most

$$\int_{|(\cdot)-q|<L/32} \frac{1}{|x-(\cdot)|} e^{-r|x-(\cdot)|} |d\chi_{\diamond q}|^2 (|\hat{h}_{pq\perp}|^2 + \Sigma_{a=0,1,2,3} z_a^2 |o_a|^2) \ .$$

(5.43)

The contribution to this integral from $|\hat{h}_{pq\perp}|^2$ is at most $c_0 L^{-2} r^{-1} \|\hat{h}_{pq\perp}\|_6^2$. Meanwhile, the $L^6$ norm of $\hat{h}_{pq\perp}$ is no greater than $c_0 \|\hat{h}_{pq\perp}\|_{\mathbb{H}}$, this being a standard Sobolev inequality because if $q \in \mathbb{H}$, then $\|q\|_{\mathbb{H}} \leq \|d|q|\|_2$. Since the contribution to (5.43) from $|\hat{h}_{pq\perp}|^2$ is at most $c_0 L^{-2} r^{-1} \|\hat{h}_{pq\perp}\|_{\mathbb{H}}^2$, the bound in (5.12) implies that it is less than $c_0 (Lr)^{-1} \sup_{x \in V_q} |\hat{h}_{p\infty}|^2$.

Let $o \in \{o_a\}_{a=0,1,2,3}$ and let $z$ denote the corresponding $z_a$. It follows from what is said in Part 1 of Section 5d that $|o| \leq c_0 L^{-2}$ on the support of $|d\chi_{\diamond q}|$ and it follows from what is said in Part 2 of Section 5d that $|z| \leq c_0 L^2 \sup_{x \in V_q} |\hat{h}_{p\infty}|$. These bounds lead to a $c_0 (rL)^{-2} \sup_{x \in V_q} |\hat{h}_{p\infty}|^2$ bound for the $z^2 |o|^2$ contribution to (5.43); this being no greater than $c_0 (rL)^{-2} (z_L^2 + z_T^2) (\frac{L}{|p-q|+L})^4$.

The conclusions of the preceding two paragraphs with (3.29) and the definitions in (5.18) and (5.19) lead to a $c_0 m^{-1/2} (\ln N)^{-1} (z_L^2 + z_T^2) (\frac{L}{|p-q|+L})^4$ bound for (5.43).

*Part 4*: The contribution to the right most integral in (5.39) from the radius $\frac{3}{16}$ ball centered on a given $q \in \Theta - p$ is at most

$$\int_{|(\cdot)-q|<3L/16} \frac{1}{|x-(\cdot)|^2} e^{-r|x-(\cdot)|} |\hat{h}_{p\infty T}| (|\mathfrak{F}_{qL}||\hat{h}_{p\infty T}| + |\mathfrak{F}_{qT}||\hat{h}_{p\infty L}| + |d\chi_{\diamond q}||\hat{h}_{pq}|) \ .$$

(5.44)

The argument in Part 3 that that leads from (5.40) to (5.41) in the case when $|x-q| > \frac{3}{16} L$ can be repeated with only cosmetic changes to prove that (5.41) also bounds (5.44) when $|x-q| > \frac{3}{16} L$. Likewise, the same argument that was used in Part 3 to derive (5.42) can be used to bound the contribution to (5.44) from $|\hat{h}_{p\infty T}|(|\mathfrak{F}_{qL}||\hat{h}_{p\infty T}| + |\mathfrak{F}_{qT}||\hat{h}_{p\infty L}|)$ in the case when $|x-q| \leq \frac{3}{16} L$. Meanwhile, it follows from the definition in (5.19) of $z_T$ that the contribution of $|\hat{h}_{p\infty T}|(|d\chi_{\diamond q}||\hat{h}_{pq}|)$ to (5.44) when $|x-q| \leq \frac{3}{16} L$ is no less than

$$c_0 z_T (\frac{L}{|p-q|+L})^2 \int_{|(\cdot)-q|<L/32} \frac{1}{|x-(\cdot)|^2} e^{-r|x-(\cdot)|} |d\chi_{\diamond q}||\hat{h}_{pq}| \ .$$

(5.45)

The integral factor in (5.45) is at most $c_0 (rL)^{-1/2} (z_L + z_T) (\frac{L}{|p-q|+L})^2$, this being a consequence of (5.12) given the definitions of $z_L$ and $z_T$ in (5.18) and (5.19). The latter bound leads to a $c_0 (rL)^{-1/2} z_T (z_L + z_T) (\frac{L}{|p-q|+L})^4$ bound on the whole of (5.45).

*Part 5*: This part of the proof considers the contribution to the integrals on the right hand side (5.39) from the the part of $\mathbb{R}^3$ where the distance to $\Theta-p$ is greater than $\frac{3}{16}L$. Consider first the contribution to the left most integral in (5.39) from the radius $\frac{1}{4}L$ ball centered at p when $|x-p| > \frac{1}{4}L$. Use (5.2) and the fact that $|w_p| \le 1$ to bound this contribution by

$$c_0 \, (m^{-3} N^2 L^{-2} + L^2 t) \, e^{-r|x-p|} \, e^{\frac{1}{4}rL} \, .$$

(5.46)

Since $L^2 = m^{-3/2} N$, this is at most $c_0 \, m^{-3/2} N \, e^{-r|x-p|} \, e^{\frac{1}{4}rL}$. When $|x-p| \le \frac{1}{4}L$, the bound in (5.2) and the $|w_p| \le 1$ bound lead to a $c_0 \, m^{-3/2} N$ bound for the contribution to the left most integral in (5.39) from the radius $\frac{1}{4}L$ ball centered at p.

The contribution to the left most integral on the right hand side of (5.39) from the part of $\mathbb{R}^3$ where the distance to $\Theta-p$ is greater than $\frac{3}{16}L$ and the distance to p is greater than $\frac{1}{4}L$ is at most

$$c_0 \int_{L/4 < |(\cdot)-p|} \frac{1}{|x-(\cdot)|} e^{-\sqrt{\frac{5}{4}}r|x-(\cdot)|} e^{-r|(\cdot)-p|} e^{-\frac{1}{4}rL} \, .$$

(5.47)

This integral can be computed in closed form with the result being that it is no greater than $c_0 \, r^{-2} \, e^{-r|x-p|} \, e^{\frac{1}{4}rL}$ when $|x-p| > \frac{1}{4}L$ and less than $c_0 \, r^{-2}$ when $|x-p| \le \frac{1}{4}L$.

Consider next the contribution to the right most integral in (5.39) from the radius $\frac{1}{4}L$ ball centered at p in the case when $|x-p| > \frac{1}{4}L$. It follows from the definition of $z_T$ in (5.19) that this contribution is at most

$$c_0 z_T e^{-r|x-p|} e^{\frac{1}{4}rL} \int_{|(\cdot)-p|<L/4} \frac{1}{|x-(\cdot)|^2} (|d\tau_p||\hat{h}_p| + \tau_p |w_p|)$$

(5.48)

The bound in (5.3) and the $|w_p| \le 1$ bound lead to a $c_0 (L^{-1/2} m^{-1} N^{3/4} (\ln N)^{1/2} + L)$ bound for the integral factor in (5.48). Since $L = m^{-3/4} N^{1/2}$ it follows from (3.28) that the contribution to the left most integral in (5.39) from the radius $\frac{1}{4}L$ ball centered at p is no greater than $c_0 z_T \, m^{-5/8} N^{1/2} (\ln N)^{1/2} e^{-r|x-p|} e^{\frac{1}{4}rL}$ when $|x-p| > \frac{1}{4}L$.

When $|x-p| \le \frac{1}{4}L$, the contribution to the the right most integral in (5.39) from the radius $\frac{1}{4}L$ ball centered at p when $|x-p| \le \frac{1}{4}$ is bounded by $c_0 z_T$ times the integral factor in (5.48) and thus by $c_0 z_T \, m^{-5/8} N^{1/2} (\ln N)^{1/2}$.

The contribution to the right most integral in (5.39) from the part of $\mathbb{R}^3$ where the distance to $\Theta-q$ is greater than $\frac{3}{16}L$ and the distance to p is greater than $\frac{1}{4}L$ has $k = w_p$ in the integrand. This being the case, it follows from the definition of $z_T$ in (5.19) that this contribution is no larger than

$$c_0 z_T \int_{L/4 < |(\cdot)-p|} \frac{1}{|x-(\cdot)|^2} e^{-\sqrt{\frac{5}{4}} r |x-(\cdot)|} e^{-r|(\cdot)-p|} e^{-\frac{1}{4}rL} .$$

(5.49)

This integral is no greater than $c_0 z_T r^{-1} e^{-r|x-p|} e^{\frac{1}{4}rL}$ in the case when $|x-p| \geq \frac{1}{4}L$ and no greater than $c_0 r^{-1}$ when $|x-p| \leq \frac{1}{4}L$.

*Part 6*: If $x \in \mathbb{R}^3$ has distance greater than $\frac{3}{16}L$ from each $q \in \Theta-p$ and distance greater than $\frac{1}{4}L$ from p, then the bounds from Parts 3-5 for the right hand side of (5.39) bound $|\hat{h}_{p\infty T}|$ at x by

$$\tfrac{1}{100}(z_T + z_L)(e^{-\frac{1}{2}r|x-p|} e^{\frac{1}{8}rL} + \sum_{q\in\Theta-p} e^{-\frac{1}{2}r|x-q|} e^{\frac{3}{32}rL} (\tfrac{L}{|p-q|+L})^2) + $$
$$c_0 m^{-5/8} N^{1/2} (\ln N)^{1/2} e^{-\frac{1}{2}r|x-p|} e^{\frac{1}{8}rL} .$$

(5.50)

On the other hand, if $q \in \Theta-p$ and x has distance less than $\frac{3}{16}L$ from q, then what is said in Parts 3-5 bound $|\hat{h}_{p\infty T}|$ at x by

$$\tfrac{1}{100}(z_T + z_L)(\tfrac{L}{|p-q|+L})^2 + c_0 m^{-5/8} N^{1/2} (\ln N)^{1/2} e^{-\frac{1}{2}r|p-q|} e^{\frac{1}{8}rL} .$$

(5.51)

Meanwhile, if x has distance less than $\frac{1}{4}L$ from p, then the bounds from Parts 3-5 lead for the right hand side of (5.39) lead to a bound for $|\hat{h}_{p\infty T}|$ at x by

$$\tfrac{1}{100}(z_T + z_L) + c_0 m^{-5/8} N^{1/2} (\ln N)^{1/2} .$$

(5.52)

It follows from the definition of $z_T$ in (5.19) that there exists a point in $\mathbb{R}^3$, this denoted by x, where one of the following conditions holds:

- *The distance from x to $\Theta-p$ is greater than $\frac{3}{16}L$, the distance to p is greater than $\frac{1}{4}L$, and $|\hat{h}_{p\infty T}|$ at x is greater than $\frac{9}{10} z_T (e^{-\frac{1}{2}r|x-p|} e^{\frac{1}{8}rL} + \sum_{q\in\Theta-p} (\tfrac{L}{|p-q|+L})^2 e^{-\frac{1}{2}r|x-q|} e^{\frac{3}{32}rL}).$*

- *The distance from* x *to a point* q ∈ Θ−p *is less than* $\frac{3}{16}$L *and* $|\hat{h}_{p\infty T}|$ *at* x *is greater than* $\frac{9}{10} z_T (\frac{L}{|p-q| + L})^2$.
- *The distance form* x *to* p *is less than* $\frac{1}{4}$L *and* $|\hat{h}_{p\infty T}|$ *is greater than* $\frac{9}{10} z_T$.

(5.53)

Use (5.50) if the top bullet in (5.53) describes x, use (5.51) if the middle bullet describes x, and use (5.52) if the third bullet in (5.53) describes x. The relevant bound for $|\hat{h}_{p\infty T}|$ leads directly to the bound in (5.34).

**References**

[Bol]   S. Bolognesi, *Multi-monopoles and magnetic bags*, Nucl. Phys. **B752** (2006) 93-123.

[EG]    J. Evslin and S. B. Gudnason, *High Q monopole bags are urchins*, preprint Nov. 2011 (arXiv:1111.3891).

[HPS]   D. Harland, S. Palmer and C. Sämann, *Magnetic domains*, preprint Apr. 2012 (arXiv:1204.6685v1).

[JT]    A. Jaffe and C. H. Taubes, <u>Vortices and Monopoles</u>, Birkhauser Boston 1980.

[LW]    K.-M. Lee and E. J. Weinberg, *BPS magnetic monopole bags*, Phys. Rev. **D79** (2009) 025013

[M]     N. Manton, *Monopole planets and galaxies*, Phys. Rev. D., to appear. Preprint Nov. 2011 (arXiv:1111.2934v2).

[PS]    M. K. Prasad and C. M. Sommerfield, *Exact classical solution for the 't Hooft monopole and the Julia-Zee dyon*, Phys. Rev. Lett. **35** (1975) 760-762.

[R]     H. L. Roydent, <u>Real analysis</u>, Prentice Hall 1988.

[RSZ]   E. A. Rakhmanov, E. B. Saff and Y. M Zhou, *Electrons on the sphere*, in <u>Computational Methods and Function Theory</u> (R. M. Ali, St. Ruscheweyh and E. B. Saff, eds.), World Scientific (1995) 111-127.

[S1]    D. Singleton, *Exact Schwarzchild-like solutions for Yang-Mills theories'*, Phys. Rev. **D51** (1995) 5911-5914.

[S2]    D. Singleton, *General relativistic analog solutions for Yang-Mills theory'*, Theor. Math. Phys. **117** (1998) 1351-1363.

[SK]    A. B. J. Kuijlaars and E. B. Saff, *Distributing many points on a sphere*, The Mathematical Intelligencer, **19** (1997) 5-11.

[T]     C. H. Taubes, *A gauge invariant index theorem for asymptotically flat manifolds*, in <u>Asymptotic behavior of mass and spacetime geometry</u>, Lecture Notes in Phys. **202** (1984) 85-94.